\documentclass[apj]{emulateapj}
\def\icarus{\ref@jnl{Icarus}} 

\usepackage{amssymb,amsmath,latexsym,natbib,wasysym, mathtools}

\usepackage{xcolor}
\usepackage[hang,flushmargin]{footmisc} 
\usepackage{url}
\makeatletter
\g@addto@macro{\UrlBreaks}{\UrlOrds}
\makeatother

\DeclareMathAlphabet{\mathpzc}{OT1}{pzc}{m}{it}

\newcommand\be{\begin{equation}}
\newcommand\ee{\end{equation}}

\newcommand{\ik}{{\it Kepler~}}
\newcommand{\kstarns}{Kepler-80}
\newcommand{\kstar}{Kepler-80~}

\newcommand{\mys}{Kepler-80f}
\newcommand{\myt}{Kepler-80d}
\newcommand{\myu}{Kepler-80e}
\newcommand{\myv}{Kepler-80b}
\newcommand{\myw}{Kepler-80c}

\def\kms{\ifmmode{\rm km\thinspace s^{-1}}\else km\thinspace s$^{-1}$\fi}
\def\koi{Kepler-80}

\def\x{\times}
\begin{document}
%\tracingall
% ^this^ is what breaks it 

\pagenumbering{arabic}

\slugcomment{Submitted to Astrophysical Journal}
\shorttitle{Five Planets of \kstarns}
\shortauthors{MacDonald et al.}

\title{A Dynamical Analysis of the Kepler-80 System of Five Transiting Planets}

\author{Mariah G. MacDonald\altaffilmark{1}, Darin Ragozzine\altaffilmark{1,2,3}, Daniel C. Fabrycky\altaffilmark{4}, Eric B. Ford\altaffilmark{5}, Matthew J. Holman\altaffilmark{3}, Howard T. Isaacson\altaffilmark{6}, Jack J. Lissauer\altaffilmark{7}, Eric D. Lopez\altaffilmark{8}, Tsevi Mazeh\altaffilmark{9}, Leslie Rogers\altaffilmark{4}, Jason F. Rowe\altaffilmark{10,11}, Jason H. Steffen\altaffilmark{12}, Guillermo Torres\altaffilmark{3}}

\email{mmacdonald2012@my.fit.edu, darin.ragozzine@gmail.com}

\altaffiltext{1}{Florida Institute of Technology, Department of Physics and Space Sciences, 150 West University Blvd., Melbourne, FL 32940, USA}
\altaffiltext{2}{University of Florida, Department of Astronomy, 211 Bryant Space Science Center, Gainesville, FL 32611, USA}
\altaffiltext{3}{Harvard-Smithsonian Center for Astrophysics, 60 Garden St., Cambridge, MA 02138, USA}
\altaffiltext{4}{Department of Astronomy and Astrophysics, University of Chicago, 5640 South Ellis Avenue, Chicago, IL 60637, USA}
\altaffiltext{5}{Department of Astronomy \& Astrophysics, and Center for Exoplanets and Habitable Worlds, 525 Davey Lab, The Pennsylvania State
University, University Park, PA, 16802, USA}
\altaffiltext{6}{Department of Astronomy, University of California, Berkeley, Berkeley, California 94720, USA}
\altaffiltext{7}{Space Science and Astrobiology Division, MS 245-3, NASA Ames Research Center, Moffett Field, CA 94035, USA}
\altaffiltext{8}{Institute for Astronomy, Royal Observatory Edinburgh, University of Edinburgh, Blackford Hill, Edinburgh, UK}
\altaffiltext{9}{School of Physics and Astronomy, Raymond and Beverly Sackler Faculty of Exact Sciences, Tel Aviv University, Tel Aviv 69978, Israel}
\altaffiltext{10}{NASA Ames Research Center, Moffett Field, CA 94035, USA}
\altaffiltext{11}{SETI Institute, Mountain View, CA 94043, USA}
\altaffiltext{12}{University of Nevada, Las Vegas, Department of Physics and Astronomy, 4505 S. Maryland Parkway, Box 454002, Las Vegas, NV 89154-4002, USA}

\setcounter{footnote}{0}

\begin{abstract} 

\ik has discovered hundreds of systems with multiple transiting exoplanets which hold tremendous potential both individually and collectively for understanding the formation and evolution of planetary systems. Many of these systems consist of multiple small planets with periods less than $\sim$50 days known as Systems with Tightly-spaced Inner Planets, or STIPs. One especially intriguing STIP, Kepler-80 (KOI-500), contains five transiting planets: f, d, e, b, and c with periods of 1.0, 3.1, 4.6, 7.1, 9.5 days, respectively. We provide measurements of transit times and a transit timing variation (TTV) dynamical analysis. We find that TTVs cannot reliably detect eccentricities for this system, though mass estimates are not affected. Restricting the eccentricity to a reasonable range, we infer masses for the outer four planets (d, e, b, and c) to be $6.75^{+0.69}_{-0.51}$, $4.13^{+0.81}_{-0.95}$, $6.93^{+1.05}_{-0.70}$, and $6.74^{+1.23}_{-0.86}$ Earth masses, respectively. The similar masses but different radii are consistent with terrestrial compositions for d and e and $\sim$2\% H/He envelopes for b and c. We confirm that the outer four planets are in a rare dynamical configuration with four interconnected three-body resonances that are librating with few degree amplitudes. We present a formation model that can reproduce the observed configuration by starting with a multi-resonant chain and introducing dissipation. Overall, the information-rich Kepler-80 planets provide an important perspective into exoplanetary systems. 

\end{abstract}


%\maketitle

\section{Introduction}
\setcounter{footnote}{0}

\ik has solidified the existence of a new population of planetary systems that consist of multiple small (1-3 Earth radii), nearly coplanar planets with periods 
concentrated around 5-50 days \citep[e.g.,][]{2011ApJ...736...19B, 2011ApJS..197....8L} now known as STIPs or Systems with Tightly-spaced Inner Planets (see $\S$\ref{stips}). Though the earliest examples were discovered with radial velocity surveys \citep[e.g.,][]{2006Natur.441..305L,2011arXiv1109.2497M}, \emph{Kepler} has significantly expanded our understanding of this population with its discovery of hundreds of stars with multiple transiting planet candidates \citep[e.g.][]{2011ApJ...736...19B, 2011ApJS..197....8L,2015arXiv151206149C}. 

Technically, some of the multi-transiting systems are composed of planet  candidates, and some of these candidates might be false positives. While $Kepler$'s false 
positive rate is generally low due to careful candidate vetting, it is clear that candidates in systems with multiple \ik 
candidates are much more likely to be planets \citep{RH10, 2010ApJ...713L.140L,2011ApJS..197....8L}, especially those with 3 or more candidates \citep{2012ApJ...750..112L,2014ApJ...784...45R,2014ApJ...784...44L}, whose purity is near 99\%. This purity is only one example of the value of multi-transiting systems, with many more aspects discussed in \citet{RH10} and subsequent works. 

Complementary to studies of the multi-transiting systems as an ensemble are investigations into individual systems to infer the masses of the planets from their mutual gravitational interactions as manifested in deviations from a perfectly periodic sequence of transits. These non-Keplerian motions are characterized by measuring how the times of transits are non-periodic, whence the now-common name of Transit Timing Variations (TTVs). While the value of TTVs was predicted before \ik \citep[e.g.,][]{2005MNRAS.359..567A, 2005Sci...307.1288H}, \ik has measured hundreds  of statistically significant TTVs \citep{2013ApJS..208...16M,2015arXiv150400707R}, allowing for precise and numerous mass estimates, particularly of small planets that are difficult to detect with Radial Velocity (RV) measurements \citep[see, e.g., ][]{2014ApJS..210...20M,2014PNAS..11112616F, 2015ARA&A..53..409W,2016MNRAS.tmp...32S}. TTVs in multi-transiting systems are particularly valuable, since the combination of masses and radii can yield multiple density measurements in a single system. 

In this work, we present such a detailed study for the transiting planets of \kstar (also KOI-500, KIC 4852528, 2MASS J19442701+3958436). \kstar has the historical distinction of being the first system identified with 5-candidates. The outer two candidates in this system were confirmed by observing anti-correlated transit timing variations and were called Kepler-80b and Kepler-80c with periods of 7.1 and 9.5 days, respectively \citep{2013ApJS..208...22X}. The middle two candidates were validated by \citet{2014ApJ...784...44L} and \citet{2014ApJ...784...45R} and were named Kepler-80d and Kepler-80e with periods of 3.1 and 4.6 days respectively. \citet{2016ApJ...822...86M} recently validated the innermost 1.0-day period planet, now Kepler-80f. In this work, we are able to measure the masses of the outer four planets, identify their dynamical relationship, and simulate their formation. 

We present the observations and data ($\S$\ref{obs and data}) and the inferred  stellar properties ($\S$\ref{stellar props}). We then turn to a detailed analysis of the TTV data for \kstar ($\S$\ref{ttv}), including a validation of our fitting procedure and assumptions ($\S$\ref{validation}). With mass estimates, we investigate the physical properties of the planets, including the mass fraction of H/He gas ($\S$\ref{planet props}). We then explore the dynamical configuration of Kepler-80, finding its planets to be in multiple three-body resonances ($\S$\ref{dynamics}). A simulation showing the formation of the system that achieves the observed three-body resonant configuration is presented in $\S$\ref{formation}. Finally, we summarize our conclusions and look forward to future observational and theoretical investigations ($\S$\ref{conclusion}). 

%%%%%%%%%%%%%%%%%%%%%%%%%%%%%%%%%%%%%%%%%%%%%%%%%%%%%%%%%%%%%%%%%%%%%%%%%%%%%%%%%%%%

%%%%%%%%%%%%%%%%%%%%%%%%%%%%%%%%%%%%%%%%%%%%%%%%%%%%%%%%%%%%%%%%%%%%%%%%%%%%%%%%%%%%

\section{Observations and Data} \label{obs and data}

\subsection{\ik Photometry} \label{photometry}

\kstar was observed photometrically by the \emph{Kepler Space Telescope} and is subject to the benefits and limitations of this method as described in numerous publications\footnote{Many relevant publications can be found at  \texttt{http://keplerscience.arc.nasa.gov/data-products.html} and \texttt{http://archive.stsci.edu/kepler/data\char`_release.html}}. 
\kstar fell on Module 3, which suffered a failure early in the \ik Mission and which resulted in the loss of data from Quarters 6, 10, and 14. \kstar also has the historical distinction of being the first system identified with 5-candidates (although superseded even at that time by the 6-planet Kepler-11 reported in \citet{2011Natur.470...53L}); there are now $\sim$20 such systems. This early detection is consistent with the relatively high SNR of each planet and the high confidence that the signals are truly due to planetary transits and not some kind of false positive or false alarm \citep{2015arXiv151206149C}. \citet{2014ApJ...784...44L} displays the folded light curves of the Kepler-80 planets in their Figure 10. Due to its early detection, the \emph{Kepler TTV/Multi-planets Working Group} recommended \kstar for short cadence observations which were obtained in Quarters 7,8,9,11,12,13,15,16, and 17. 

We had access to several sets of Transit Timing (TT) measurements, including the publicly available data from \citet{2015arXiv150400707R} and \citet{2013ApJS..208...16M}. We also had the updated long cadence TT estimates from the Mazeh group (Holczer et al. 2016, submitted) and short cadence TT data from both co-authors JR and DF. These were all measured using similar methods \citep[see][]{2013ApJS..208...16M} and had no major differences. 

We fit TT data from some of these sources with our full dynamical model, and again generally received consistent results. As our goal was to identify the optimal dataset for getting the highest precision masses, we compared all these data directly by examining the scatter of TT measurements after subtracting the best-fit quadratic+sinusoid model, which is a good approximation of the overall TTV model. The short cadence TT data from co-author DF showed somewhat lower scatter than the other data, so our final results are based on these data, which we report in Table \ref{table:scdata}. Fits of different datasets were investigated, and we do not think that these would give statistically significant inconsistencies in the final planetary properties. We also note here that, although the Kepler-80 system is in a Laplace-like three-body resonance, its planets are too small and have not been observed long enough to see the TTV trends expected for such systems by \citet{2013MNRAS.430.1369L}.

The TT data and uncertainties used for the main fit (Table \ref{table:scdata}) were generated by optimizing each individual transit time. In particular, the data (\texttt{SAP\_FLUX}) are divided by an \texttt{occultsmall} transit shape \citep{2002ApJ...580L.171M} and the residuals fit to a polynomial to implement detrending.  All the transits have the same shape but different transit mid-times and uncertainties. If the center of a transit fell within 500 minutes of the center of a transit of Kepler-80b or c (the only two that have significant individual transits), then that transit was not used in the analysis, nor its TT reported. 

As in \citet{2016ApJ...820...39J}, we found that the residuals to our quadratic+sinusoid models were much better approximated by a Student t-distribution with 2 degrees of freedom than by a Gaussian distribution. This distribution is indicative of ``heavy tails'', e.g., a statistically significant excess of large residuals compared to a Gaussian distribution. This motivated the use of a non-Gaussian (``$t_2$'') error model, as discussed below. Another benefit of this error model is that it is robust to TT outliers. Obvious outliers were identified using a visual inspection of the lightcurve, but given this robustness, we elected not to remove any potential outliers from the TT measurements. 

\begin{deluxetable*}{llcccc}
\tablecaption{Short Cadence Transit Time Data\label{table:scdata}}
\tablecolumns{6}
\tablewidth{0pt}
\tabletypesize{\small}
\tablehead{\colhead{KOI} & \colhead{Kepler Name} & \colhead{Transit No.} & \colhead{Transit Time} & \colhead{TTV}& \colhead{TT Error}}   
\startdata

500.03  & Kepler-80d &    -236 &        70.095100 &      0.0170892 & 
    0.00924170 \\
500.03  &Kepler-80d&    -235 &        73.167282 &     -0.0143836 & 
    0.00925570 \\
500.03  &Kepler-80d&    -234 &        76.239464 &    -0.00131950 & 
    0.00936270 \\
500.03  &Kepler-80d&    -233 &        79.311645 &     -0.0197255 & 
     0.0162559 \\
500.03  &Kepler-80d&    -232 &        82.383827 &     0.00371700 & 
    0.00962480 \\
500.03  &Kepler-80d&    -231 &        85.456009 &     -0.0110081 & 
    0.00868490 \\
500.03  &Kepler-80d&    -227 &        97.744743 &     0.00282010 & 
    0.00819760 \\
    
\hline

\enddata
\tablecomments{Short cadence data used for the TTV fitting, reduced by author DF. The columns, from left to right, are: the planet's KOI number, the planet's Kepler name, the transit number (where transit 0 indicates the first transit after the epoch of 793), the transit time (BJD - 2454900), and the transit timing variation, and the uncertainty in the transit time. All times are in units of days. Table \ref{table:scdata} is published in its entirety in the electronic edition of the \emph{Astrophysical Journal}. A portion is shown here for guidance regarding its form and content.}
\end{deluxetable*}

\subsection{Spectroscopy}

Spectra were taken of \kstar by Keck and McDonald Observatories, and these spectra and preliminary interpretations are available on the Kepler Community Follow-up Observing Program (CFOP) website\footnote{\texttt{https://cfop.ipac.caltech.edu}}. We acquired an 1800 second high resolution spectrum with the Keck I telescope and the HIRES spectrometer on 2011 July 20. The standard California Planet Search setup and data reduction of HIRES \citep{2009ApJ...696...75H} was used, resulting  in a  SNR of 35 at 5500 Angstroms.  The C2 decker, with dimensions of 0.87" x 14", was used to allow a resolution of $\sim$60,000 and sky subtraction. Sky subtraction is required to produce reliable spectra for stars as faint as \kstarns\ ($K_p = 14.8$). This spectrum was the primary source for stellar classification. 

\citet{2016arXiv160408604B} observed Kepler-80 with medium robo-AO quality and found no companions, and we assume that there is no significant contamination of stellar or planetary properties from additional stars. Although future investigation may reveal blending or dilution, for our analysis in this paper, we assume that these are negligible. 

\section{Stellar Properties}\label{stellar props}

The stellar properties of \koi\ have been somewhat elusive, as
different techniques originally gave different answers. In particular,
\cite{2012ApJ...750L..37M} included \koi\ in their analyses of M dwarfs
observed by {\it Kepler} and found an effective temperature near
4000~K based on near-infrared spectroscopy, corresponding to a spectral type of M0; however, additional
investigation into that result, as well as new spectral and photometric
analyses described below, present a clear story that motivates our
adopted stellar classification as a K5 dwarf.

A common complication in some methods of analysis of optical spectra
is the strong correlation between the derived effective temperature
($T_{\rm eff}$), metallicity ([Fe/H]), and surface gravity ($\log g$).
This correlation can prevent the determination of accurate values of these three
quantities simultaneously \citep[see, e.g.,][]{2012ApJ...757..161T}. In order to
obtain an independent estimate of the temperature, we turned to
available standard photometry of \koi.  Brightness measurements in the
Sloan system ($griz$) were corrected for known zero-point offsets
\citep[see][]{2012ApJS..199...30P} and transformed to the Johnson-Cousins
system using some 40 different published relations \citep{2005AN....326..321B,
  2008MNRAS.384.1178B, 2011MNRAS.417.2230B, 2005AJ....130..873J, 2006A&A...460..339J, 2006AJ....132..989R, 2006PASP..118.1679D, 2008AJ....135..264C} as well as unpublished ones by
Lupton~(2005)\footnote{\url{https://www.sdss3.org/dr8/algorithms/sdssUBVRITransform.php\#Lupton2005}},
of which some involve also the WISE (W1, W2) and 2MASS ($JHK_s$)
magnitudes. We then used all of these relations to solve for best
compromise values of the $BV(RI)_{\rm C}$ magnitudes, obtaining $B =
16.342 \pm 0.079$, $V = 15.188 \pm 0.057$, $R_{\rm C} = 14.442 \pm
0.069$, and $I_{\rm C} = 13.833 \pm 0.071$. With these results and the
2MASS magnitudes, we constructed eight different but non-independent
color indices, and used the calibrations of \cite{2010A&A...512A..54C} to
infer effective temperatures. Solar composition was assumed for the
metallicity terms in these calibrations, although this is a very small
effect ($< 10$~K).  Reddening was estimated using the prescriptions by
\cite{1998ApJ...500..525S}, \cite{2003A&A...409..205D}, \cite{2005AJ....130..659A}, and
\cite{2015ApJ...810...25G}, adopting a preliminary distance estimate of
360~pc. The mean of the four consistent values, $E(B-V) = 0.068 \pm
0.020$, was applied to de-redden the color indices mentioned above
prior to computing the temperatures. The corresponding extinction to
\koi\, assuming $A_V = 3.1 E(B-V)$, is $A_V = 0.21 \pm 0.06$.  The
weighted average temperature we obtained from the eight color indices
is $T_{\rm eff} = 4530 \pm 100$~K, corresponding to spectral type K5.

To place an indirect constraint on $\log g$, we proceeded as
follows. First, we made use of the measured rotation period of the star of 25.6 days from \citet{2014ApJS..211...24M}
and \citet{2015ApJ...800..142M}. This rotation period is manifested clearly as a $\sim$0.5\% amplitude variation in the raw (\texttt{SAP\_FLUX}) photometry and is presumably due to starspots. The rotation period was combined  with the gyrochronology relations of
\cite{2007ApJ...669.1167B}, \cite{2008ApJ...687.1264M}, \cite{2009ApJ...695..679M},
\cite{2010ApJ...722..222B}, and \cite{2014ApJ...780..159E} to infer an age for the
system. Values ranged from 1.3~Gyr to 2.9~Gyr with a mean of about
2~Gyr, to which we assigned an uncertainty of 1~Gyr so as to encompass
the lowest and highest estimates. We then used the temperature derived
above and this age, along with a solar-metallicity model isochrone
from the Dartmouth stellar evolution series \citep{2008ApJS..178...89D}, to
obtain a crude surface gravity estimate of $\log g \approx 4.6$. This
value was then adopted for our spectroscopic analysis of the
HIRES/Keck~I spectrum, using the Spectroscopy Made Easy methodology
(SME). This analysis resulted in values of $T_{\rm
  eff} = 4540 \pm 88$~K, ${\rm [Fe/H]} = +0.04 \pm 0.08$, and an upper
limit on the projected rotational velocity $v \sin i$ of $1 \pm
1$~\kms.

Given the excellent agreement between the spectroscopic and
photometric temperatures, we proceeded to a more detailed comparison
with the Dartmouth models, using the spectroscopic $T_{\rm eff}$ value
(with an uncertainty conservatively increased from 88~K to 100~K), the
corresponding metallicity, and the age derived above. This age
estimate serves as a good proxy for luminosity or $\log g$, given that
isochrones in the $\log g$ vs.\ $T_{\rm eff}$ diagram are essentially
horizontal at this temperature. Our Monte Carlo procedure for
comparing the observations with the models resulted in an estimated
stellar mass of $M = 0.730 \pm 0.030~M_{\odot}$, a radius of $R = 0.678
\pm 0.023~R_{\odot}$, a mean stellar density of $\rho = 2.33 \pm
0.15~\rho_{\odot}$, and a bolometric luminosity of $L =
0.170_{-0.022}^{+0.033}~L_{\odot}$.  Inferred absolute magnitudes in
the $V$ and $K_s$ bands are $M_V = 7.24 \pm 0.25$ and $M_{K_s} = 4.47
\pm 0.11$. The resulting surface gravity from this fit, $\log g =
4.639 \pm 0.012$ (cgs), is sufficiently close to the value adopted for
the SME analysis that no iteration is necessary.  As a further
consistency check, we used a 2~Gyr Dartmouth isochrone for the measured
metallicity to solve simultaneously for the distance and reddening
values that provide the best fit to the Sloan and 2MASS photometry. We
obtained $E(B-V) = 0.060$, in good agreement with our previous
estimate, and a distance of $D \approx 357$~pc. We adopt these stellar
parameters (Table \ref{table:stellarprops}) for the remainder of this paper.

An additional reduction of the Keck HIRES spectrum on the CFOP website also finds consistent results (Sam Quinn, pers. comm.). An additional spectrum with lower quality was taken by the MacDonald spectrum, which also led to a consistent conclusion. 

Why then did \citet{2012ApJ...750L..37M} claim an effective temperature of 4000 K? Their methodology used infrared spectroscopy to classify low mass stars. This technique, applied to stars warmer than $\sim$4000 K, can lead to some misinterpretation, and, accounting for this systematic error, the true error bar from \citet{2012ApJ...750L..37M} should be $\sim$250 K (P. Muirhead, personal communication). Therefore, a warmer star is actually consistent with all of the data gathered and is the solution we adopt for Kepler-80.  

Some of our planetary properties are derived from the Markov Chain Monte Carlo (MCMC) results of \citet{2014ApJ...784...45R} located on the ExoplanetArchive \texttt{http://exoplanetarchive.ipac.caltech.edu/}. These chains result in a temperature of $4613\pm74$ K, log $g$ of $4.690\pm0.06$, stellar density of $2.8\pm0.2$ $\rho_{\odot}$, stellar radius of $0.637\pm0.022$ $R_{\odot}$, and mass of $0.72\pm0.11$ $M_{\odot}$. These properties are mostly consistent with the more detailed spectroscopic method. When determining planet parameters, we choose to combine some of the MCMC results with our spectroscopic stellar parameters; for example, we take the planet to star radius ratios from the MCMC analysis and combine them with our stellar radius estimate to estimate the planetary radii with uncertainties (see $\S$\ref{ttv} for full details). Though this combination is not entirely self-consistent, it is not a major concern due to the similarity of the stellar parameters inferred by the two methods. 

\begin{deluxetable*}{lllll}
\tablecaption{Stellar Properties of Kepler-80\label{table:stellarprops}}
\tablecolumns{6}
\tablewidth{0pt}
\tabletypesize{\footnotesize}
\tablehead{\colhead{Parameter} & \colhead{Value} & \colhead{$1\sigma$ Error}&\colhead{Units}}   
\startdata

Kepler ID (KIC/KID) & 4852528 & & \\
Right Ascension & 19:44:27.02  & & hh:mm:ss   \\
Declination &39:58:43.6   & & dd:mm:ss \\
Kepler magnitude ($K_p$) & 14.804 & \nodata & mag \\
Spectral Type & K5 & \nodata & \nodata \\
Distance & 357 & \nodata & pc  \\
Effective Temperature ($T_{eff}$) & 4540 & 100 & K  \\
Surface Gravity (log $g$) & 4.639 & 0.012 & [cgs]  \\
Metallicity ([Fe/H]) & 0.04 & 0.08 & \nodata  \\
Radius ($R_{\star}$) & 0.678 & 0.023 & $R_{\odot}$ \\ 
Mass ($M_{\star}$) & 0.730 & 0.030 & $M_{\odot}$  \\
Density ($\rho_{\star}$) & 2.33 & 0.15 & g~cm$^{-3}$ \\
Luminosity ($L$) & 0.170 & $^{+0.033}_{-0.022}$ & $L_{\odot}$  \\
Absolute Magnitude ($M_V$) & 7.24 & 0.25 & mag \\
Absolute Magnitude ($M_K$) & 4.47 & 0.11 & mag \\

\enddata
\end{deluxetable*}
 
%%%%%%%%%%%%%%%%%%%%%%%%%%%%%%%%%%%%%%%%%%%%%%%%%%%%%%%%%%%%%%%%%%%%%%%%%%%%%%%%%%%%

\section{TTV analysis} \label{ttv}

\subsection{Introduction and Methods}

The strength and character of transit timing variations (TTVs) can be used to determine the masses and orbital properties, primarily of the perturbing planet(s). The outer four planets of \kstar show statistically significant TTVs with a character similar to many other Kepler TTV systems: anti-correlated sinusoids with a ``super-period'' equal to the time it takes for the line of conjunctions to circulate by one full revolution (if the planets were massless). This  super-period is also a measure of distance from the $j+1:j$ mean-motion resonance, and is given by:
\begin{equation}\label{eq:superperiod}
\frac{1}{|(j+1)/P' - j/P|} 
\end{equation}
where $P$ and $P'>P$ are the orbital periods of the two planets \citep{2005MNRAS.359..567A,2012ApJ...761..122L}. In the case of Kepler-80, the super-period for each of the three neighboring pairs 
of the outer four planets is $\sim191$ days. That multiple pairs share the same super-period is a special feature of the \kstar system, equivalent to the multiple three-body resonance configuration discussed below. 

As \mys\ is dynamically decoupled from the outer four planets, we did not include it in our TTV analysis. We also assume that the TTVs are not affected by any potential non-transiting planets; the final self-consistent fit argues against additional planets, but we did not test this explicitly. 

Our TTV model is generated using a five-body integration calculated with a Burlisch-Stoer algorithm that is optimized to determine the times, impact parameters, and velocities at the moment when the sky-projected center of the planet is closest to the center of the star. The parameters used to generate the model include, for each planet: the epoch ($T_0$), the period ($P$),  the eccentricity multiplied by the cosine and sine of the argument of periapsis ($e \cos \omega$ and $e \sin \omega$), the sky-plane inclination ($i$), the longitude of ascending node ($\Omega$), and the planet-to-star mass ratio, for a maximum of 28 parameters. The coordinate system and definitions follow the conventions in \citet{2010exop.book..217F}. 

The times from the integration are correlated to the associated times from the data and ($TTV_{model} - TTV_{observed}$) is calculated. As mentioned above, the distribution of these residuals was not Gaussian and included significant outliers. With motivation from \citet{2016ApJ...820...39J}, we elected to use a Student's $t$-distribution with 2 degrees of freedom. That is, our fit determines a likelihood by assuming that the observations are described by the proposed model plus a random error from a $t_2$ distribution. The deviations between the model and the observations were therefore scaled by the associated $t_2$ distance (e.g., the $t$-score instead of the usual Gaussian  $z$-score) and then squared. Therefore, the likelihood (or goodness-of-fit) parameter is not $\chi^2$, which assumes Gaussian errors, and we refer to it as $\Sigma t_2^2$. The maximum likelihood is obtained at the minimum value of $\Sigma t_2^2$ \citep{2016ApJ...820...39J}.  
We performed maximum likelihood fits to the data using the Levenberg-Marquardt (LM) algorithm \emph{mpfit} \citep{2009ASPC..411..251M}, a local minimization routine. We performed thousands of LM minimizations from initial conditions chosen randomly in a region of parameter space much wider than the final error bars, as in \citet{2009AJ....137.4766R}. To assist the optimization routine in proceeding from an initial guess to the global minimum, we would begin by fitting only the periods and epochs, keeping all other parameters fixed. As discussed below, a wide variety of techniques were used to understand the properties of the data and the fitting methodology. In most cases, only a portion of the 28 parameters were allowed to take on any value (``float'') sometimes within a restricted range, while the other parameters are held fixed. 

There is not enough information in the TTV signal to uniquely determine all of the parameters, a problem that has been seen in many previous TTV studies. Therefore, we consider the simplest non-trivial fits with circular coplanar orbits in $\S$\ref{circfit}, followed by fits with restricted eccentricity ranges in $\S$\ref{reccfit}. The rationale for using simpler models is discussed in $\S$\ref{validation}. In the simpler models, we must assume values for the parameters that are not fit (e.g., coplanar orbits); we discuss the evidence that these assumptions do not significantly affect the mass estimates in $\S$\ref{validation}. 

\subsection{Circular Coplanar Fits}\label{circfit}

\begin{figure*}
\centering
\includegraphics[width=.8\textwidth]{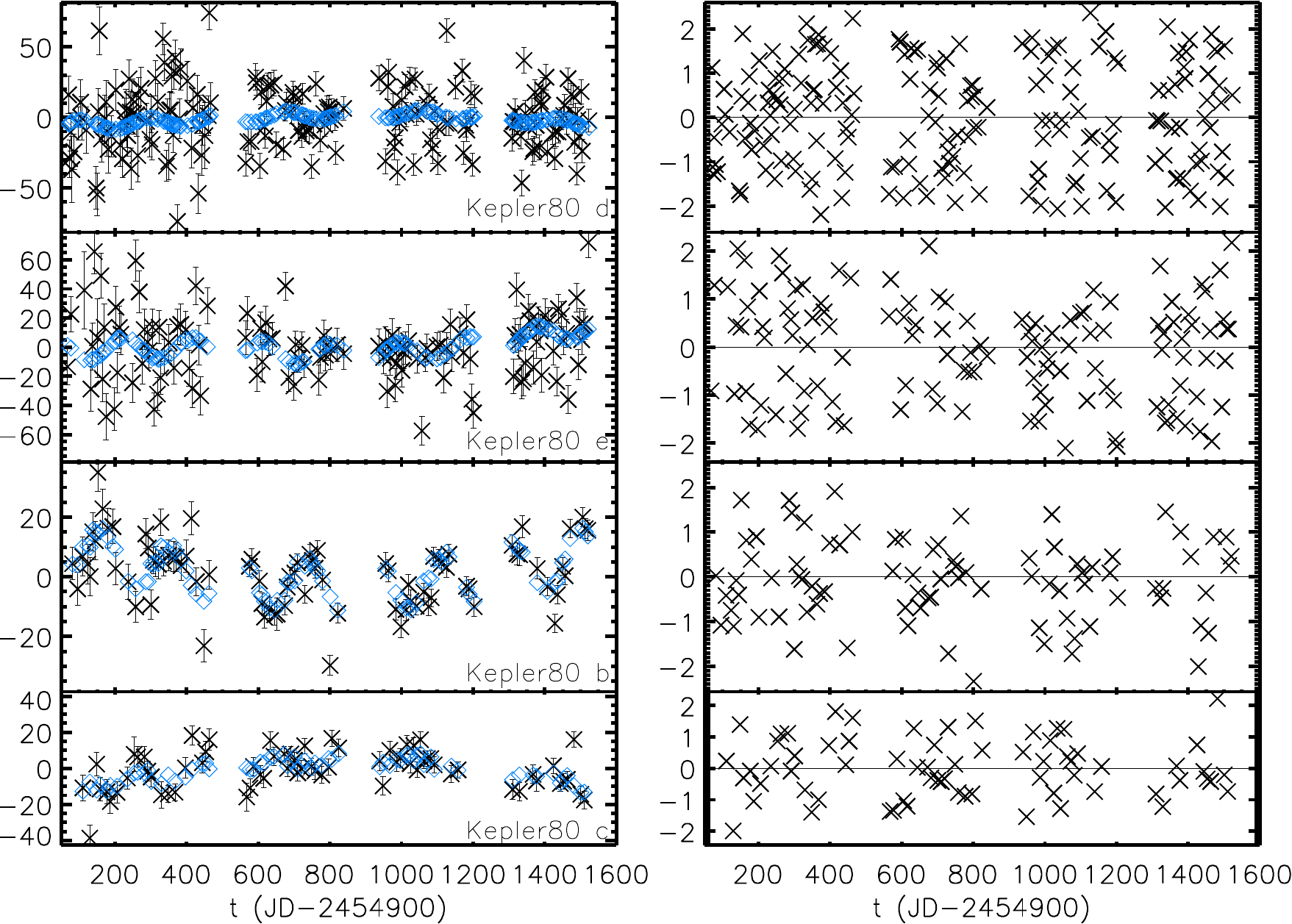}
\caption{Results of our TTV Fits. Left: the TTV data (from Table \ref{table:scdata}) is shown in black crosses with error bars. Our best-fit circular coplanar model is shown in blue diamonds. The vertical axis is Transit Time Variations from the best-fit linear ephemeris in minutes; note the varying scales. The quadratic + 191-day sinusoid nature of the model and the observations is clearly visible, particularly for the planets e, b, and c. The sinusoid is caused by the 191-day conjunction cycle and the quadratic trend is due to the $\sim$10 year libration of the three-body resonances ($\S$\ref{dynamics}). Right: the associated $t$-score of each measurement, taken by scaling the residuals (model - data) using a student t-distribution with 2 degrees of freedom, as discussed in the text. The vertical axis would correspond to the residual in units of $\sigma$ for a Gaussian distribution.\label{fig:ttvs}}
\end{figure*}

We begin by exploring the properties of circular coplanar fits, where $e \cos \omega$ and $e \sin \omega$ are fixed to 0, the sky-plane inclination is fixed to 90$^{\circ}$, and $\Omega$ is fixed to 0$^{\circ}$ for all planets. Even under these assumptions, we find excellent fits to the data, with the lowest $\Sigma t_2^2$ divided by the number of degrees of freedom (analogous to the reduced $\chi^2$) of 1.13. This best-fit circular model is shown in Figure \ref{fig:ttvs}. The model matches the 191-day sinusoid, which is caused by the cycle of planetary close approaches (conjunctions) moving all the way around the orbital plane. It also matches the quadratic trend, which is a component of the sinusoidal variation caused by the $\sim$10 year libration in the three-body resonance discussed in $\S$\ref{dynamics} below.  

Within each model, we estimate our mass uncertainties using the method of bootstrapping. We generated thirty datasets by randomly selecting, with replacement, the TTV data and performed a global minimization on each set, following the methodology above, inspecting each fit to make sure that it appeared to converge to very near the global minimum. The distributions of parameters from these bootstrapping fits are used to determine the uncertainties in our parameters. Although not a Bayesian analysis, the parameters returned from each bootstrap fit have some of the same properties as draws from a \emph{posterior} distribution; for example, we can use these to visualize the covariance between the fit parameters. We note that the mass uncertainties inferred from this bootstrap analysis are also consistent with the curvature of the $\Sigma t_2^2$ vs. parameter plots and with the uncertainties returned from Levenberg-Marquardt (which is a measurement of curvature very near the best-fit). This agreement gives us confidence that our uncertainties are well estimated.

In actuality, the TTV analysis only measures mass ratios, and the planetary mass uncertainties must be combined with the $\sim$4\% uncertainty in the stellar mass. In order to self-consistently propagate errors from stellar and planetary parameters, we combined all these results as follows. We pulled randomly with replacement from the MCMC chains of \citet{2014ApJ...784...45R} to determine the distribution of star-planet radius ratios. Keeping in mind that these were based on poorer stellar properties, a stellar mass and radius were separately drawn from a Gaussian distribution based on their adopted values in $\S$\ref{stellar props}. Similarly and independently, we pulled randomly from the distribution of the bestfit models from the thirty circular bootstrapping runs for the planet to star mass ratio, the period, and epoch. With all the stellar and planetary parameters so defined for a particular draw, we calculated the planetary mass, planetary radius, inclination, and the semi-major axis. As suggested by \citet{2014ApJ...785...15J}, the planetary density ($\rho_p$) was scaled from the stellar density ($\rho_{\star}$) using 

\begin{equation}
\label{eq:dens}
\rho_p = \rho_{\star} \left( \frac{M_p}{M_{\star}} \right) \left( \frac{R_p}{R_{\star}} \right)^3
\end{equation}

where $p$ and $\star$ refer to planetary and stellar properties. In total, we performed 1,000 draws and the results are reported in Table \ref{table:MCMCcircchains}. In Table \ref{table:MCMCresults}, we report the median and the $\pm$ 68\% confidence range for all the planetary parameters of interest for both the circular and restricted eccentricity fit. 

As expected based on the trends seen in the TTV data, all four masses are recovered with high statistical significance. Despite the different radii between the inner two planets (d and e) and the outer two planets (b and c), all the masses are similar, between 4-6 Earth masses. We discuss the implications of our mass estimates in subsequent sections. 

%%results from the bootstrap estimate for the masses are: \myt~:$	6.48^{+	0.46	}_{-	0.39	}$, %%\myu~
%%$	4.92^{+	0.49	}_{-	0.37	}$,\myv~: 
%%$	5.99^{+	0.49	}_{-	0.57	}$, and \myw~:
%%$	5.03^{+	0.39	}_{-	0.42	}$. These distributions are illustrated in Figure \ref{fig:masses}. 

\subsection{Restricted Eccentricity Fits}\label{reccfit} 

Several studies have shown that TTVs for systems like Kepler-80 near first-order mean motion resonances show a mass-eccentricity degeneracy \citep[e.g.,][]{2012ApJ...761..122L, 2014ApJ...787...80H, 2016ApJ...820...39J}. By fixing the eccentricities to be zero, we likely are underestimating the mass uncertainties. $\S$\ref{validation} below describes why allowing the eccentricities to be completely unrestricted leads to inaccurate results. In this section, we choose a compromise (similar to other TTV analyses) by restricting our fits to the low eccentricity regime. 

Specifically, we constrained $e \cos \omega$ and $e \sin \omega$ to be less than 0.02 for all planets. Otherwise, the analysis proceeded as in $\S$\ref{circfit} above including a fit to 30 bootstrapped datasets and error propagation to include uncertainties in stellar parameters. The masses are generally consistent with our circular model, though with larger uncertainties, as expected:  
$6.75^{+0.69}_{-0.51}$ for \myt, 
$4.13^{+0.81}_{-0.95}$ for \myu, 
$6.93^{+1.05}_{-0.70}$ for \myv, and
$6.74^{+1.23}_{-0.86}$ for \myw, all in units of $M_{\oplus}$. These results are presented in Table \ref{table:MCMCeccchains}.

\begin{deluxetable*}{ccccccccccccc}
\tabletypesize{\footnotesize}
\tablecolumns{13}
\tablewidth{0pt}
\tablecaption{Circular Error Propagation Analysis\label{table:MCMCcircchains}}
\tablehead{
\colhead{\#} &
\colhead{Planet} & 
\colhead{$R_{\star}$} &
\colhead{$M_{\star}$} &
\colhead{$R_p$} & 
\colhead{$M_p$} & 
\colhead{$\rho_p$} & 
\colhead{$a$} & 
\colhead{\emph{i}} &
\colhead{\emph{e}} &
\colhead{$\omega$} &
\colhead{$P$} &
\colhead{$t_0$} \\
\colhead{} &
\colhead{} & 
\colhead{$R_{\odot}$} &
\colhead{$M_{\odot}$} &
\colhead{$R_{\oplus}$} & 
\colhead{$M_{\oplus}$} & 
\colhead{g~cm$^{-3}$} & 
\colhead{AU} & 
\colhead{deg} &
\colhead{} &
\colhead{deg} &
\colhead{days} &
\colhead{days}
}

\startdata
1	&	Kepler-80d	&	0.676	&	0.768	&	1.480	&	6.675	&	7.92	&	0.038	&	89.46	&	0.00	&	0.00	&	3.0721159	&	795.12836	\\
2	&	Kepler-80d	&	0.691	&	0.730	&	1.790	&	6.520	&	4.35	&	0.037	&	85.78	&	0.00	&	0.00	&	3.0721149	&	795.12793	\\
3	&	Kepler-80d	&	0.669	&	0.808	&	1.582	&	6.268	&	5.32	&	0.039	&	86.73	&	0.00	&	0.00	&	3.072125	&	795.12897	\\
4	&	Kepler-80d	&	0.675	&	0.724	&	1.492	&	6.972	&	7.57	&	0.037	&	87.84	&	0.00	&	0.00	&	3.0721381	&	795.13031	\\
5	&	Kepler-80d	&	0.683	&	0.822	&	1.610	&	7.466	&	5.39	&	0.039	&	86.72	&	0.00	&	0.00	&	3.0721209	&	795.13037	\\
6	&	Kepler-80d	&	0.658	&	0.708	&	1.544	&	6.563	&	6.24	&	0.037	&	86.86	&	0.00	&	0.00	&	3.0721231	&	795.12915	\\
7	&	Kepler-80d	&	0.692	&	0.751	&	1.624	&	6.545	&	5.72	&	0.038	&	86.79	&	0.00	&	0.00	&	3.072098	&	795.13043	\\
8	&	Kepler-80d	&	0.666	&	0.764	&	1.469	&	6.639	&	6.91	&	0.038	&	89.11	&	0.00	&	0.00	&	3.0721159	&	795.12836	\\
9	&	Kepler-80d	&	0.702	&	0.744	&	1.701	&	6.459	&	6.01	&	0.037	&	86.45	&	0.00	&	0.00	&	3.0721159	&	795.12787	\\
10	&	Kepler-80d	&	0.676	&	0.759	&	1.464	&	6.433	&	6.97	&	0.038	&	89.90	&	0.00	&	0.00	&	3.0721231	&	795.12738	\\

\enddata
\tablecomments{In order to combine different sources of uncertainty, we employ a Monte Carlo like error propagation analysis as discussed in the main text. We performed a total of 10,000 random draws for each planet and each row represents one draw, indicated by the draw number (\#). Each draw takes a stellar mass ($M_{\star}$) and radius ($R_{\star}$) from a normal distribution based on our assumed stellar parameters (Table \ref{table:stellarprops}). The mass ratio, period ($P$), and epoch ($t_0$), eccentricity ($e$) and argument of periapsis ($\omega$) are drawn independently from the best-fits of thirty circular bootstrapping runs. Finally, a third independent draw (with replacement) is taken from the MCMC posteriors of \citet{2014ApJ...784...45R} to determine the distribution of planet-star radius ratio and impact parameter. The combination of these properties allows us to derive the planet's mass ($M_p$), period ($R_p$), density ($\rho_p$, from Equation \ref{eq:dens}), semi-major axis ($a$, from Kepler's Third Law), and sky-plane inclination ($i$). The units for each of these quantities are indicated in the second header row. Table \ref{table:MCMCcircchains} is published in its entirety in the electronic edition of the \emph{Astrophysical Journal}. A portion is shown here for guidance regarding its form and content.}
\end{deluxetable*}

\begin{deluxetable*}{ccccccccccccc}
\tabletypesize{\footnotesize}
\tablecolumns{13}
\tablewidth{0pt}
\tablecaption{Eccentric Error Propagation Analysis}
\tablehead{
\colhead{\#} &
\colhead{Planet} & 
\colhead{$R_{\star}$} &
\colhead{$M_{\star}$} &
\colhead{$R_p$} & 
\colhead{$M_p$} & 
\colhead{$\rho_p$} & 
\colhead{$a$} & 
\colhead{\emph{i}} &
\colhead{\emph{e}} &
\colhead{$\omega$} &
\colhead{$P$} &
\colhead{$t_0$} \\
\colhead{} &
\colhead{} & 
\colhead{$R_{\odot}$} &
\colhead{$M_{\odot}$} &
\colhead{$R_{\oplus}$} & 
\colhead{$M_{\oplus}$} & 
\colhead{g~cm$^{-3}$} & 
\colhead{AU} & 
\colhead{deg} &
\colhead{} &
\colhead{deg} &
\colhead{days} &
\colhead{days}
}

\startdata
1	&	Kepler-80d	&	0.676	&	0.768	&	1.480	&	7.938	&	9.42	&	0.038	&	89.46	&	0.013	&	0.00	&	3.0722229	&	795.12952	\\
2	&	Kepler-80d	&	0.691	&	0.730	&	1.790	&	6.839	&	4.57	&	0.037	&	85.78	&	0.016	&	86.25	&	3.0722439	&	795.12866	\\
3	&	Kepler-80d	&	0.669	&	0.808	&	1.582	&	6.905	&	5.86	&	0.039	&	86.73	&	0.013	&	66.20	&	3.072217	&	795.13092	\\
4	&	Kepler-80d	&	0.675	&	0.724	&	1.492	&	6.133	&	6.66	&	0.037	&	87.84	&	0.018	&	57.12	&	3.0722511	&	795.13116	\\
5	&	Kepler-80d	&	0.683	&	0.822	&	1.610	&	8.449	&	6.11	&	0.039	&	86.72	&	0.015	&	68.19	&	3.0722649	&	795.1311	\\
6	&	Kepler-80d	&	0.658	&	0.708	&	1.544	&	6.576	&	6.25	&	0.037	&	86.86	&	0.017	&	45.12	&	3.0722921	&	795.12946	\\
7	&	Kepler-80d	&	0.692	&	0.751	&	1.624	&	7.325	&	6.40	&	0.038	&	86.79	&	0.015	&	61.31	&	3.072186	&	795.13129	\\
8	&	Kepler-80d	&	0.666	&	0.764	&	1.469	&	7.895	&	8.21	&	0.038	&	89.11	&	0.013	&	48.33	&	3.0722229	&	795.12952	\\
9	&	Kepler-80d	&	0.702	&	0.744	&	1.701	&	7.074	&	6.58	&	0.037	&	86.45	&	0.003	&	52.00	&	3.0721869	&	795.12909	\\
10	&	Kepler-80d	&	0.676	&	0.759	&	1.464	&	7.053	&	7.65	&	0.038	&	89.90	&	0.015	&	92.20	&	3.0722139	&	795.12689	\\

\enddata
\tablecomments{Table columns have the same meaning as in Table \ref{table:MCMCcircchains}. Note that the values of eccentricity ($e$) and argument of periapse ($\omega$) are probably inaccurate based on the discussion in $\S$\ref{validation}. Table \ref{table:MCMCeccchains} is published in its entirety in the electronic edition of the \emph{Astrophysical Journal}. A portion is shown here for guidance regarding its form and content.\label{table:MCMCeccchains}}
\end{deluxetable*}

We view these restricted eccentricity results as the most appropriate and generally adopt these values for additional analysis, subject to the caveats described in $\S$\ref{validation} and elsewhere. In particular, we do not think that the eccentricity and periapse angle are reliably inferred from these fits, but for completeness and reproducibility, we include the recovered values in Table \ref{table:MCMCeccchains}. 

\begin{deluxetable*}{lccccc}
\tabletypesize{\scriptsize}
\tablecolumns{6}
\tablewidth{0pt}
\tablecaption{Results from Error Propagation Analysis}
\tablehead{
\colhead{Parameter} &
\colhead{Kepler-80f} &
\colhead{Kepler-80d} &
\colhead{Kepler-80e} &
\colhead{Kepler-80b} &
\colhead{Kepler-80c}
}

\startdata

Radius ($R_{\oplus}$) 	&	$1.21^{+0.06}_{-0.05}$	&	 $	1.53	^{+	0.09	}_{-	0.07	}$ 	&	 $	1.60	^{+	0.08	}_{-	0.07	}$ 	&	 $	2.67 \pm 0.10$ 	&	 $	2.74	^{+	0.12	}_{-	0.10	}$ \\
Mass, Ecc ($M_{\oplus}$) 	&	\nodata	&	 $    6.75 ^{+	0.69	}_{-	0.51	}$ 	&	 $4.13 ^{+	0.81	}_{-	0.95	}$ 	&	 $6.93 ^{+	1.05	}_{-	0.70	}$ 	&	 $6.74	^{+	1.23	}_{-	0.86	} $ \\
Mass, Circ($M_{\oplus}$) 	&	\nodata	&	 $	6.48	^{+	0.46	}_{-	0.39	}$ 	&	 $	4.92	^{+	0.49	}_{-	0.37	}$ 	&	 $	5.99	^{+	0.49	}_{-	0.57	}$ 	&	 $	5.03	^{+	0.40	}_{-	0.42	}$ \\
Density, Ecc (g~cm$^{-3}$) 	&	\nodata	&	 $	7.04 \pm1.06 $ 	&	 $	3.75	^{+	0.89	}_{-	0.97	}$ 	&	 $	1.38	^{+	0.24	}_{-	0.17	}$ 	&	 $	1.22	^{+	0.23	}_{-	0.18	}$ \\
Density, Circ (g~cm$^{-3}$) 	&	\nodata	&	 $	6.73	^{+	0.83	}_{-	0.97	}$ 	&	 $	4.54 \pm0.67$ 	&	 $	1.19	^{+	0.14	}_{-	0.13	}$ 	&	 $	0.91 \pm 0.11$ \\
Semi-major axis (AU) 	&	$0.0175\pm0.0002$	&	 $	0.0372	\pm 0.0005$ 	&	 $	0.0491 \pm 0.0007$ 	&	 $	0.0648 \pm 0.0009$ 	&	 $	0.0792 \pm0.0011$ \\
Inclination (deg) 	&	$86.50^{+2.36}_{-2.59}$	&	 $	88.35	^{+	1.12	}_{-	1.51	}$ 	&	  $	88.79	^{+	0.84	}_{-	1.07	}$ 	&	  $	89.34	^{+	0.46	}_{-	0.62	}$ 	&	 $	89.33	^{+	0.47	}_{-	0.57	}$ \\
Period (days) 	&	$0.9867873\pm0.00000006$	&	 $	3.07222	^{+	0.00006	}_{-	0.00004	}$ 	&	 $	4.64489	^{+	0.00020	}_{-	0.00019	}$ 	&	 $	7.05246	^{+	0.00020	}_{-	0.00022	}$ 	&	 $	9.52355	^{+	0.00041	}_{-	0.00029	}$ \\
Epoch (days, BJD-2454900) 	&	\nodata	&	 $	795.129	^{+	0.002	}_{-	0.001	}$  	&	$	796.915	\pm 0.002 $  	&	 $	758.399	^{+	0.002	}_{-	0.001	}$  	&	 $	796.047	\pm 0.001$  \\

\enddata
\tablecomments{Summary of the results from the 10,000 draws from the error propagation analysis given in Tables \ref{table:MCMCcircchains} and \ref{table:MCMCeccchains}. The nominal value for each parameter is taken from the median of the distributions from the error propagation analysis, and lower and upper uncertainties are taken to include the 16$^{th}$ and 84$^{th}$ percentile confidence intervals. The rows, from top to bottom, are: the planetary radius ($R_p$), the planetary mass derived from our restricted eccentricity (``Ecc'') fit ($M_p$), the planetary mass derived from our circular (``Circ'') fit, the planetary density ($\rho_p$) derived from the restricted eccentricity fit, the planetary density derived from the circular fit, the semi-major axis ($a$), the sky-plane inclination ($i$), the period ($P$) and the epoch ($t_0$, BJD - 2454900). The units for each parameter are given in parentheses. The results for the radius, the semi-major axis, the inclination are identical for the two different fits (since they do not depend on the TTV analysis) while the period and epoch are practically identical. We prefer the restricted eccentricity solution for reasons described in the text. In addition, since we did not include \mys\ in the TTV fitting (it is dynamically decoupled), values for its mass, density, and epoch cannot be included. Assuming Earth-like composition, the mass for \mys\ would be $\sim$1.8$\pm0.3 M_{\oplus}$. \label{table:MCMCresults}}

\end{deluxetable*}

\begin{figure*}[t]
\centering
\includegraphics[width=.95\textwidth]{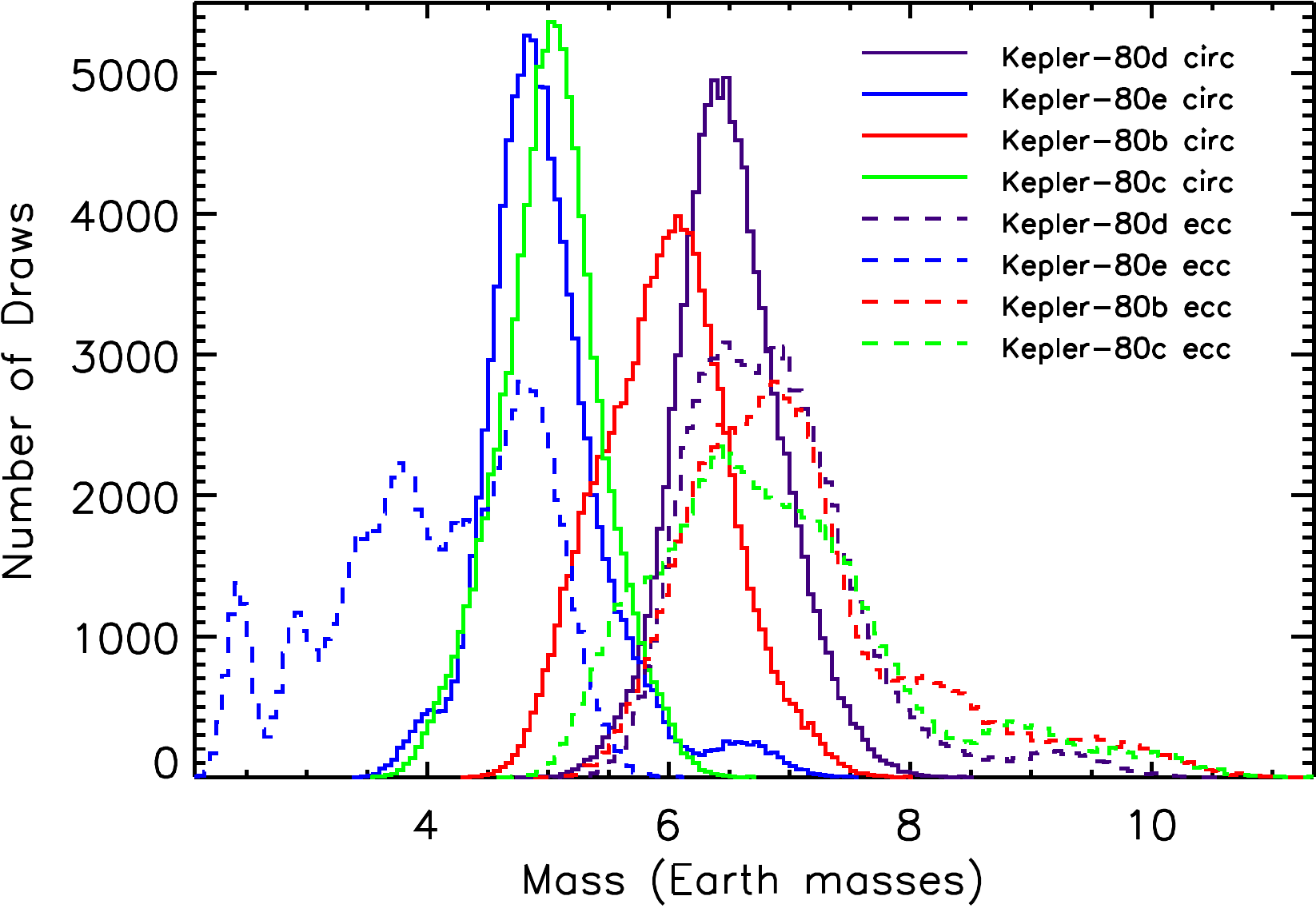}
\caption{Mass estimates for all four planets for the circular and restricted eccentricity fits. These histograms represent the frequency distribution of planetary masses (in Earth masses) for 100,000 draws from our error propagation analysis described in $\S$\ref{circfit}, which combines uncertainties in the mass ratio fits from a bootstrap analysis with stellar parameter uncertainties. Different colors correspond to different planets (d - purple, e - blue, b - red, c - green), listed in order of increasing period. Solid lines correspond to fits assuming circular orbits and have narrower distributions than fits that allowed for a restricted eccentricity range, shown with dashed lines. All four planets have similar masses and are all very reliably detected by the TTV analysis. \label{fig:masses}}
\end{figure*}

A graphical representation of the mass estimates and uncertainties of the four planets for the circular and restricted eccentricity fits is given in Figure \ref{fig:masses}. Aside from the narrower distribution (smaller uncertainty) for the circular fits, the lowest mass planet e (4.6-day period) also shows a multi-modal mass distribution in the eccentric case, with the circular fit occupying only one of the modes. There is also a discrepancy between the circular and restricted eccentricity fits for the outermost planet b (9.5-day period) which is not statistically significant, though a bit worrisome. Based on the analysis in $\S$\ref{validation}, we did not expect major differences. 

\begin{figure}
    \centering
    \includegraphics[width=.45\textwidth]{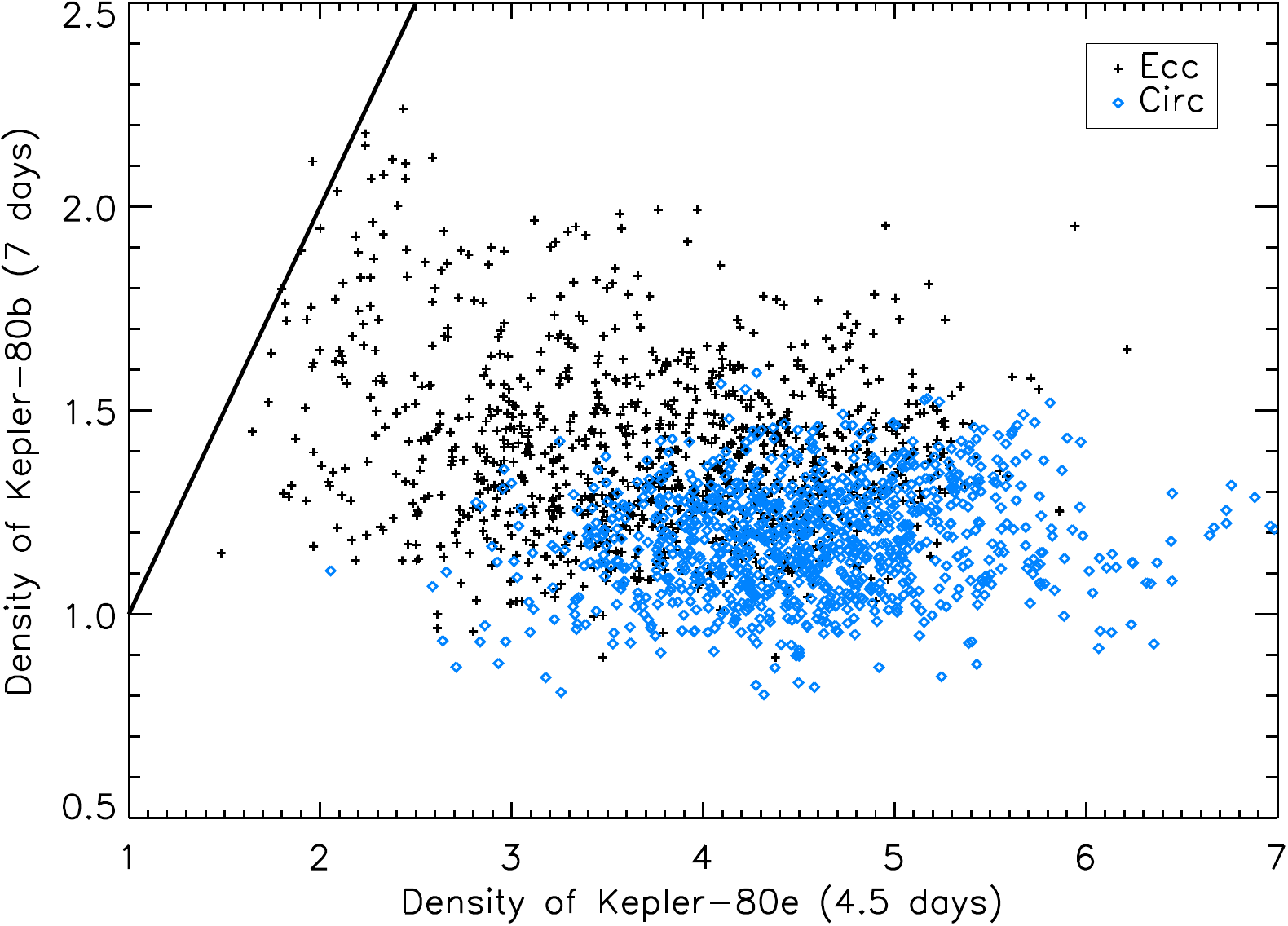}
    \caption{Neighboring planets Kepler-80e and Kepler-80b have very different densities (in units of g~cm$^{-3}$). These differences are seen in both the circular and the restricted eccentricity fits. Note that the axes are on different scales, which is illustrated by the solid 1:1 line that would indicate equal densities. Kepler-80b clearly has a lower density than Kepler-80e, even when including the larger uncertainty from the restricted eccentricity fit.\label{fig:densityoplot}}
\end{figure}

The larger uncertainties in masses for the restricted eccentricity fits correspond to a larger uncertainty in densities; however, the very different radii between the inner two planets and the outer two planets yield densities that are clearly different. Figure \ref{fig:densityoplot} compares the density estimate for neighboring planets e (4.5-day period) and b (7-day period), including a 1:1 line that shows clearly that e is more dense than b. The implications for these density differences are discussed in subsequent sections. 

\subsection{Validation of TTV Fitting Methods} \label{validation}

We have shown that circular and restricted eccentricity fits provide clear mass estimates with reasonable uncertainties. In order to motivate and validate those analyses, we performed several additional exercises to understand the properties of fits to the Kepler-80 TTVs. 

With sufficient signal-to-noise, TTVs can be used to solve for masses and all orbital parameters (excepting the sky orientation). Practically speaking, Kepler-80 and most other systems do not have the precision necessary for a complete solution, and a different strategy is needed. Letting all parameters float in the fit is nominally the appropriate technique with subsequent pruning or interpretation to deal with unusual results. For Kepler-80, the global minimization led to highly-inclined orbits with eccentricities of $\sim$0.2, apsidally aligned into nested orbits. This configuration conflicts with long-term stability, expected properties of the system based on the ensemble of Kepler STIPs \citep[e.g.][]{2014ApJ...790..146F}, and the likelihood of seeing all the planets currently transiting \citep{2015AAS...22525728B}. To understand why such a solution is obtained, we performed a series of tests of our methodology. 

These tests generally started with planetary masses of 5.7, 2.1, 4.6, and 4.3 Earth masses based on a preliminary fit of an earlier TTV dataset from \citet{2013yCat..22080016M} where no parameters were fixed. In these preliminary fits, to compensate for outliers, we used a robust (in the statistical sense) ``truncated $\chi^2$'' error model, where the goodness-of-fit was given by removing 10\% of the highest residuals and then calculating $\chi^2$ using Gaussian errors in the normal fashion. Uncertainties on the parameters were estimated using the curvature of the parameter vs. truncated $\chi^2$ distribution from the hundreds or thousands of Levenberg-Marquardt minimizations. The rest of the fitting procedure was the same as described above. Although these tests study a slightly different method than we presented in fits above ($\S$\ref{circfit} and \ref{reccfit}), we think they are similar enough that the results translate well to our main results. 

\subsubsection{Tests Related to Eccentricities}
\label{circular}

When the eccentricities were allowed to take any value in the fitting process, the fit would invariably approach eccentricities of 0.1-0.2 and nested orbits, similar to that seen for other $Kepler$ planetary systems by \citet{2015Natur.522..321J} and \citet{2016MNRAS.455L.104G}. This solution implies that the Kepler-80 data were insufficient to fully break the mass-eccentricity degeneracy \citep[e.g.,][]{2012ApJ...761..122L, 2016ApJ...820...39J}, which is reasonable given the quality of the TTV data. 

More importantly, we found that this tendency to go to large aligned eccentric orbits was due to overfitting. Following the parameters of Kepler-80, we generated fake data (with errors taken from the real data) based on a circular model that was then fit with a model where eccentricities were allowed to float. The resulting fit strongly preferred aligned orbits with large eccentricities, just as we saw for the real data, which clearly indicates that the eccentricities derived from these data are not a property of the actual planets. We performed additional investigations (different masses, different starting conditions) along these lines to confirm that TTVs cannot reliably estimate eccentricities for the Kepler-80 system. This insight strongly motivates the use of a fitting methodology that constrains the eccentricities to a reasonable range as in our restricted eccentricity fit. This is essentially the same issue seen in other TTV studies, such as \citet{2016ApJ...820...39J}, who handle this issue using restrictive eccentricity priors in a Bayesian analysis. 

In a similar vein, we generated fake data with eccentricities that were not apsidally aligned and found that the fitting process resulted in alignment that was not present in the actual (fake) model. The approximate degeneracy between mass and eccentricity is actually a complex interplay between certain components of the eccentricity vectors at different frequencies and with different strengths which tends to produce fits with strong degeneracy along apsidal alignment \citep{2014ApJ...787...80H, 2016ApJ...820...39J}. 

We thus caution that dynamical interpretation of eccentricities and apsidal angles from the TTV fits to Kepler-80 could be severely over-interpreted. As apsidal alignment or anti-alignment is a feature of some formation simulations \citep[e.g.,][]{2016MNRAS.455L.104G}, the known degeneracies of TTV fitting must be carefully excluded to avoid drawing inaccurate conclusions. 

A similar concern arises when attempting to ascertain whether the planets in the system are librating in two-body resonances, as discussed below. 

Despite these issues, we note that other TTV analyses indicate that the mass-eccentricity degeneracy does not preclude reliable mass estimates, even when the eccentricities are not well known \citep{2015Natur.522..321J, 2016ApJ...820...39J}. To confirm these results, we performed several different parametrizations of the Kepler-80 case and explored a wide variety of ``fake'' data fitting. Even when our fake datasets led to inaccurate eccentricities and apsidal angles, the masses were recovered within 1-$\sigma$ (based on the error of each particular model) with one exception (expected given the number of tests performed) as described in Table \ref{tab:masses}. (Although our best-fit masses differ from the masses shown here, this discussion is focused on validating the methods and not the final results.) 

Our conclusion is that the eccentricities, whether small or large, whether fixed or floating, do not significantly affect the estimates of the masses. As allowing for eccentric orbits causes the global minimization to go to a known inaccurate high eccentricity state due to overfitting, we elect to focus on models with zero ($\S$\ref{circfit}) or restricted ($\S$\ref{reccfit}) eccentricities. 

\subsubsection{Tests Related to Inclinations}
\label{coplanar}
Through investigation using synthetic data sets, we determined that TTVs do not depend on Kepler-80 being nearly co-planar (see Section \ref{coplanar}). Letting the inclinations and longitudes of ascending nodes float (except for planet d, see Ragozzine \& Holman 2010) resulted in very non-coplanar fits, but with large error bars on the mutual inclinations. Similarly, we were able to readily recover accurate masses of a fake dataset that was generated with planets with 2-3$^{\circ}$ inclination, but fit with a coplanar model. This result is consistent with theoretical \citep[e.g.][]{2005MNRAS.359..567A,2012ApJ...761..122L} and empirical \citep[e.g.][]{2010ApJ...712L..86P} expectations. Based on inclinations derived from the impact parameters from the MCMC chains, it appears that some relative inclinations at the $\sim$1$^\circ$ level are possible, but we note that a detailed analysis of the lightcurves with the new stellar parameters would be required to fully justify any inference of mutual inclinations. 
 
While producing TTV models, we also produced models of Transit Duration Variations (TDVs) in this system for a wide variety of eccentric and inclined orbits. Some early fits included a chi-square penalty for incorrect transit durations and explored the regularization parameter that would be needed to combine TTV and transit duration measurements. Assuming small ($\lesssim$5$^{\circ}$) inclinations yield TDVs in this system that are near or below the threshold of detectability. The model TDVs show that the model durations primarily vary on the same 191-day timescale as the TTVs, due to short-term variability in the orbits and not due to the much longer secular precession timescale. This result holds for a wide variety of inclinations and eccentricities and is consistent with existing TDV measurements \citep[e.g.][]{2013ApJ...777....3N} which detect short-term variability and with estimates that TDV signals for secular variation are generally undetectable \citep{2015AAS...22525728B}. The claim of \citet{2009ApJ...698.1778R} that ``transit shaping'' would be much more important than ``transit timing'' is only justified when the precession timescale is sufficiently short, i.e., for very hot Jupiters, and is not applicable to the vast majority of \ik systems. 

\subsubsection{Tests Related to Masses}

All of the fits illustrated in Table \ref{tab:masses} use the same ``true'' masses in the generation of fake data. We also confirmed with additional analyses that fake datasets generated with both different masses and different mass ratios also resulted in the inference of accurate masses within uncertainties. Yet another test started the analysis with an initial guess far from the true masses and was also successful. Some of these tests allowed eccentricities to float and others forced circular orbits; in accordance with the results above, this did not make a significant difference in most cases. 

As the radii of planets d and e are smaller, the TTV data are not as clean, although the model TTV amplitudes are comparable. The expected 191-day periodicity is seen in the power spectra of the TTVs of all 4 planets. To ensure that the masses of these planets were clearly detectable with the data, we created a fake dataset where the two inner planets masses were zero and the fitter correctly recovered this result for both forced circular and floating eccentricities. Hence, we have confidence that we have reliably measured the masses of the planets d and e.

\subsection{Summary of TTV Fitting}

The assumption of coplanar and circular or near-circular orbits in our main TTV fits is required by the need to avoid overfitting that leads to inaccurate eccentricity estimates. Recognizing that reliable eccentricity estimates are not possible, we focus on retrieving accurate masses. Our tests confirm other analyses which find that mass estimates are not significantly affected by the unknown eccentricities. To avoid underestimating our mass uncertainties, we prefer using a model that allows for a restricted range of eccentricities. We again emphasize that eccentricities recovered from this model are most likely inaccurate. 

It is worth noting that this extensive testing also validates our global minimization methodology of thousands of local Levenberg-Marquardt minimizations. Investigation of the global fits on the 60 bootstrap datasets also showed reasonable convergence and recovery, giving further confidence to our analysis. We also confirmed that adopting a smaller integration timestep did not significantly affect our results. Adopting the masses and uncertainties from the restricted eccentricity model, we now explore the implications of these mass measurements for understanding the Kepler-80 system. 

\section{Planet Properties} \label{planet props}

\begin{deluxetable*}{ccrccrccrccrccrcc}
\small
\tablecaption{Testing Mass Robustness\label{tab:masses}}
\tabletypesize{\footnotesize}
\tablewidth{0pt}
\tablecolumns{17}
\tablehead{
\colhead{Pl} & \colhead{Truth}	&	\multicolumn{3}{c}{Everything} 
& \multicolumn{3}{c}{Circular} &
\multicolumn{3}{c}{Nested 1} & \multicolumn{3}{c}{Nested 2}	&  	\multicolumn{3}{c}{Inclination}	\\
& & $M_p$ & $\sigma$	& $d$& $M_p$ & $\sigma$	& $d$ & $M_p$ & $\sigma$	& $d$ & $M_p$ & $\sigma$	& $d$ & $M_p$ & $\sigma$	& $d$
}   

\startdata

d & 5.7 & 5.4 & 1.5 & -0.2 & 5.3 & 0.7 & -0.6 &	5.5	&	0.5	& -0.4 &	7.2	&	1.5	& +1.0 &	5.9	&	1.5	& +0.1 \\
e & 2.1 & 2.8 & 0.9 & +0.8 & 2.1 & 0.5 & 0.0 &	2.5	&	0.5	& +0.8 &	2.6	&	0.5	& +1.0 &	1.8	&	1.2	& -0.3	\\
b & 4.6 & 4.1 & 1.7 & -0.3 & 4.5 & 0.8 & -0.1 &	5.3	&	0.8	& +0.9 &	5.0	& 0.5 & +1.0 & 4.7 &	0.9	& +0.1 	\\
c & 4.3 & 3.5 & 1.8 & -0.4 & 4.0 & 0.5 & -0.6 &	4.5	&	0.6	& +0.3 &	4.8	& 0.8 & +0.6 & 4.3 &	0.8	& 0.0 	

\enddata
\tablecomments{Recovered masses from our investigation into fake datasets. All masses are in units of $M_{\oplus}$. Here, $M_p$ is the recovered mass, $\sigma$ is the recovered standard deviation, and $d$ is the deviation from the "truth," in units of $\sigma$. ``Everything'' stands for everything floats, meaning that no parameters were fixed or restricted. ``Circular'' stands for fits where the ``true'' orbits were circular (see $\S$\ref{circular}). ``Nested 1'' and ``Nested 2'' were runs where the $\omega$ values were fixed to the same value so that the orbits were nested inside of each other.  ``Inclination'' is the fixed inclination runs (see $\S$\ref{coplanar}). It is clear that practically all masses were returned within or at 1-$\sigma$. This table summarizes most, but not all of the fits used to validate our methodology (see$\S$\ref{validation}.}
\end{deluxetable*}

The precision in our recovered masses justifies an investigation into the physical properties of these planets. In particular, it is relatively rare to have multiple low-mass planets with well-measured densities in the same system; furthermore, the Kepler-80 planets are quite close to each other in physical distance from the star. The commonalities expected between these planets justify some discussion of comparative planetology. 

At $\sim$1.2 Earth radii and with the significant stellar insolation received in its 1-day orbit, it seems very likely that \mys\ is a rocky planet. If so, we estimate its mass by assuming that it follows a mass-radius relationship illustrated by \citet{2015ApJ...800..135D}, who found that very small planets with precise densities are consistent with an Earth-like rock-to-iron ratio. Using this relation and the uncertainty in the planetary radius and stellar properties, we estimate the mass of \mys\ to be 1.8$M_{\oplus}$ ($\pm 0.3$), but emphasize that this is a best-guess extrapolation and not a measurement. 

\begin{figure*}						 
\centering
\includegraphics[scale=0.70]{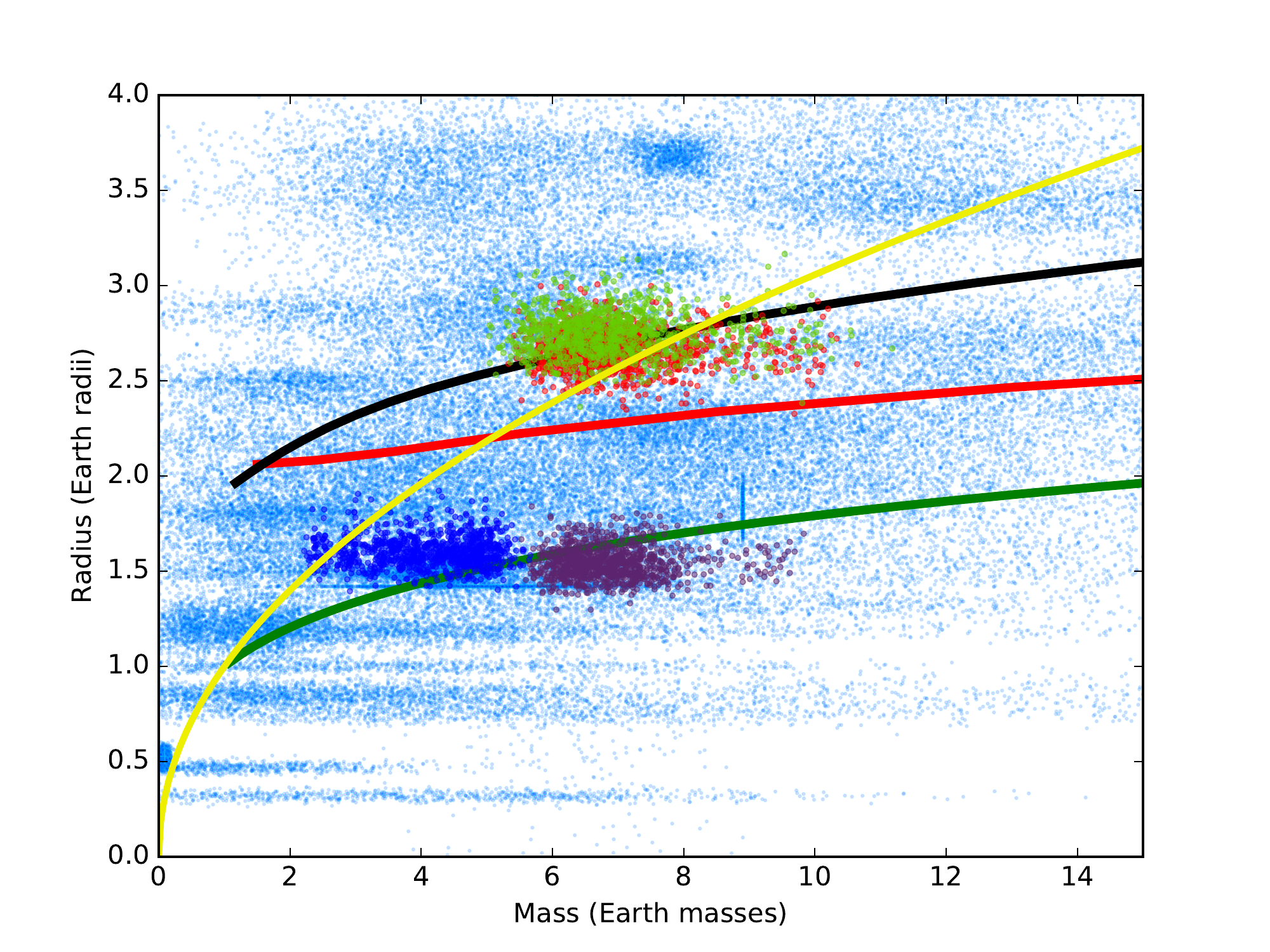}
	\caption{\label{fig:mr} Mass-Radius diagram with Kepler-80 planets. Standard plots with error bars give more visual real estate to planets with larger uncertainties. We therefore employ a new mass-radius diagram that plots 1000 semi-transparent points based on a Monte Carlo estimate of the masses and radii and their reported uncertainties for 107 planets with measured properties (most of which are off of the plot). This estimate does not account for any correlation in uncertainties, but does account for asymmetric uncertainties, if reported. The four planets of Kepler-80 are plotted as four distinct colors (80b - red, 80c - green, 80d - purple, 80e - blue) consistent with other Figures. The broad background of known planets indicates that there is no simple mass-radius relationship for small planets, both due to large uncertainties and to the underlying distribution \citep[e.g.,][]{2015arXiv150407557W}. It also shows that the Kepler-80 planets have masses and radii consistent with other planets, though with smaller uncertainties than many other estimates. The solid lines truncated near 1 Earth mass show constant composition curves from \citet{2014ApJ...792....1L} of (top to bottom) pure water worlds (black), an Earth-like core with a 1\% H/He fraction (red), and Earth-like compositions (green), respectively. The yellow solid line that begins in the origin shows the empirical mass-radius estimate of $M = R^{2.06}$ used as a rough approximation by \citet{2011ApJS..197....8L}. Though all four of Kepler-80's planets have similar masses, \myt\ and \myu\ have nearly Earth-like compositions and \myv\ while \myw\ (which overlap in the plot) can be explained by an Earth-like core beneath a $\sim$2\% H/He envelope.  }
\end{figure*}

We place the outer four planets on a mass-radius diagram focused on small planets in Figure \ref{fig:mr}. We employ a new mass-radius diagram that plots 1000 Monte Carlo realizations of the mass-radius relation based on the reported asymmetric (but uncorrelated) uncertainties of 107 masses and radii (most of which are off the range of the plot). The colored points are the new results from Kepler-80 presented here (taken from Table \ref{table:MCMCeccchains}). Although all four of Kepler-80's planets have similar masses, they separate into two groups of two planets based on radius. The inner two, \myt\ and \myu, are similar to terrestrial planets with Earth-like compositions, while the outer two (\myv~and \myw) must have some H/He envelope.

Using models based on \citet{2011ApJ...738...59R}, \citet{2010ApJ...716.1208R}, \citet{2010ApJ...712..974R} and results from the error propagation analysis shown in Table \ref{table:MCMCeccchains}, we analyze potential compositions for all four planets moving outward from the parent star. (As Kepler-80f does not have a mass measurement, we do not consider it.)

For Kepler-80d, we find that $>99\%$ of the samples are more dense than pure silicates and only 0.9\% of samples demand any volatiles. We find that 89\% of the samples are more dense than Earth-like composition. When fitting the planet with a two layer rocky-planet model, consisting of an iron core surrounded by a (Mg \#90) silicate mantle, we find an iron core mass fraction of $53^{+14}_{-19} \%$ by mass. Assuming an Earth-like composition rocky interior, we find a 95$^{\rm th}$ percentile upper limit on surface H$_2$O mass fraction of $<1\%$ or an upper limit of 0.05\% H/He by mass. This planet is clearly terrestrial in nature.

For Kepler-80e, we find that 42\% of samples are more dense than pure silicate composition, and 58\% of samples demand volatiles. When fitting the planet with a two layer rocky-planet model, consisting of an iron core surrounded by a (Mg \#90) silicate mantle, we find an iron core mass fraction of $15 ^{+15}_{-12} \%$ by mass, with a 95\% upper limit of 41\% by mass. Upon adopting an Earth-like composition rocky interior, we find a 95th percentile upper limit on surface H$_2$O mass fraction of 22\%.
Note, however, that this planet is the least understood and that the restricted eccentricity solution shows multiple mass modes. If the circular fit is correct and the largest mode is actually preferred (see Figure \ref{fig:masses}), the density of the planet is increased, lowering the need for volatiles and making this planet terrestrial in nature. 

For Kepler-80b, we find that all samples are less dense than iron-poor silicate composition, requiring volatiles. In addition, 85\% of samples are less dense than a (implausible) pure-water composition, requiring an envelope of light gasses. All but 0.01\% of samples are less dense than a more plausible 50-50 Earth-like rocky and water composition. We find a plausible composition to be $1.5^{+0.4}_{-0.3} \%$ by mass H/He atop an Earth-like composition core. In this case, the envelope accounts for the outer $\sim$38\% of the planet radius (and $\sim$76\% of the planet's volume). Similarly, Kepler-80c must have a H/He layer and one possible composition is a $1.8\pm 0.4\%$ H/He layer (by mass) atop an Earth-like composition core; the envelope then accounts for the outer $\sim$41\% of the planet radius.

Based on the models of \citet{2014ApJ...792....1L}, we can estimate what the original gas envelopes of these planets would have been like before possible photo-evaporation. The densities require that the inner two terrestrial planets must have lost any initial H/He envelopes and, accounting for photo-evaporation, planets d and e could have had initial H/He envelopes of $\sim$0.7\% and $\sim$1.1\% at 10 MYr, respectively. Nearby planets b and c could have had H/He envelopes of $\sim$3.1\% and $\sim$3.7\% at 10 MYr, respectively. Note that these outer planets --- with the same stellar history, roughly similar distances, and similar (core) masses --- probably experienced a similar amount of photoevaporation independent of our model-dependent result of $\sim$1\% H/He by mass. 

Since planets e and b are only 0.015 AU from each other, it is interesting to speculate how they ended up with such different densities (Figure \ref{fig:densityoplot}). Compositionally, they are not very different as the addition of only a small amount of H/He ($\sim$2\% by mass) is enough to explain the very different radii \citep[e.g.,][]{2014ApJ...792....1L}. Still, we consider whether this ``jump'' in gas content can be explained by theoretical planet formation models. For example, \citet{2015ApJ...811...41L} describe scaling laws for the expected gas to core ratio that planets should be able to accrete as a function of planet core mass and equilibrium temperature (among many other features like disk lifetime which should be the same for all the planets in the system). In particular, they find that for dust-free accretion that the gas to core ratio goes as $M_{core}^{1.6} T_{eq}^{-1.9}$. Using this scaling law and our best fit values for the core masses of each of the planets, we find that the present compositions can be roughly explained, within uncertainties, by the combination of differences in \emph{in situ} accretion along with evaporation. Reduction of the density uncertainty should provide tighter constraints. 

Another explanation for the different properties could be related to different formation conditions followed by migration. For example, the outer two planets could have migrated from farther out in the disk where accretion of H/He gas is easier. The dynamical configuration suggests migration, as discussed in $\S$\ref{formation} below, but we must conclude that the present data are not sufficiently precise to clarify the formation and evolution of the planetary atmospheres. 

%%%%%%%%%%%%%%%%%%%%%%%%%%%%%%%%%%%%%%%%%%%%%%%%%%%%%%%%%%%%%%%%%%%%%%%%%%%%%%%%%%%%

%%%%%%%%%%%%%%%%%%%%%%%%%%%%%%%%%

\section{Dynamics of the Kepler-80 system} \label{dynamics}
\label{stips} 
Kepler-80 belongs to the group of planetary systems known as STIPs: Systems with Tightly-spaced\footnote{STIPs were originally defined as Systems with Tightly-\emph{packed} Inner Planets by \citet{2012DPS....4420004R}. This definition can be confusing since ``packed'' can imply ``dynamically packed'' in the sense that adding intermediate planets would result in instability. Though this was not the intention, the use of ``Tightly-\emph{spaced}'' does not carry the same dynamical connotation but retains the sense that the planets in these systems are close together compared to solar system planets.} Inner Planets. A large variety of studies \citep[see reviews by][]{2014PNAS..11112616F,2015ARA&A..53..409W} have identified the key properties of these systems, though there is no set definition. The STIPs designation was created to help describe exoplanetary systems that contain multiple relatively small planets (0.5-4 Earth radii) that orbit their star, with periods between roughly 5 and 50 days. We note that ``Inner'' planets in the STIPs acronym indicates innermost, i.e., these are the planets with the shortest orbital periods. The use of ``Inner'' does not imply that there are no additional planets, although some STIPs have known planets with longer periods. In addition, ``Tightly-spaced'' is defined in comparison to the solar system; STIPs can have similar dynamical distances between the planets as planets in the solar system (e.g., separations of $\sim$20-30 Mutual Hill Radii), but the absolute distances are smaller since planets in STIPs are closer to their parent stars \citep[see, e.g.,][]{2011A&A...528A.112L}. With these clarifications, Kepler-80 clearly qualifies as a STIP and we use our mass estimates to investigate some dynamical properties of this interesting system. 

\subsection{Dynamical Stability} \label{stability}

Previous studies \citep[e.g.,][]{2011ApJS..197....8L,2014ApJ...790..146F} performed dynamical stability analyses of \ik multi-planet systems, including Kepler-80, and found that the vast majority were stable (with the rare unstable systems likely being blends). Similarly, we performed a basic stability analysis to estimate the upper limits on the eccentricities of each of the Kepler-80 planets and found that eccentricities of 0.1-0.2 are not likely to be stable. 

With the newly estimated masses, a more detailed investigation into dynamical stability is possible, but beyond the scope of this work; however, we can use analytical formulae to estimate how close the system is to stability. We perform a stability analysis very similar to \citet{2014MNRAS.445.3959Q} and use multiple analytical stability estimates. 

A common dynamical distance estimate in multi-planetary systems is the mutual Hill radius $R_{mH}$ defined as
\begin{equation}
R_{mH} \equiv \bigg( \frac{\mu_i+\mu_{i+1}}{3} \bigg)^{1/3} \bigg( \frac{a_i + a_{i+1}}{2} \bigg),
\end{equation}
where $\mu$ is the planet-to-star mass ratio \citep[e.g.][]{2011ApJS..197....8L, 2014ApJ...790..146F}. Using masses of 1.8, 6.75, 4.13, 6.93, 6.74 $M_{\oplus}$ (and circular orbits), we find that the distance between the five planets in Kepler-80 in units of their mutual Hill radii are 29.2, 11.0, 10.4, and 7.6, respectively. 
\citet{1993Icar..106..247G} determined that a coplanar system of two planets in circular orbits is Hill stable if they are separated by more than $2\sqrt{3} \approx 3.46$ mutual Hill radii, which is clearly satisfied in this case. \citet{1980AJ.....85.1122W} found that a low mass object in nearly-circular orbit is likely to experience chaos due to overlapping first-order resonances when the normalized separation is comparable to $1.5\mu^{2/7}$, but the Kepler-80 pairs are also well beyond this stability limit.

%\citet{1996Icar..119..261C}, 
\citet{2009Icar..201..381S}, and similar studies have found that long-term stability of multi-planetary systems typically required a distance of 10 mutual Hill radii. \citet{2011ApJS..197....8L} tried to enforce this limit by suggesting that long-term instability could be an issue if the sum of two consecutive mutual Hill distances was greater than 18. The outer four planets are encroaching on these limits, but as these stability criteria are mostly heuristic in nature, their tight dynamical spacing is not necessarily indicative of long-term instability. 

\citet{2015ApJ...808..120P} found a system of two planets to be dynamically stable if
\begin{equation}
\frac{a_{i+1}(1-e_{i+1})}{a_i(1+e_i)} > 2.4\big[max(\mu_i,\mu_{i+1})^{\frac{1}{3}}\big(\frac{a_{i+1}}{a_i}\big)^{\frac{1}{2}} + 1.15\big]
\end{equation}
although, as with all these analytic criteria, the stability boundary is fuzzy. Applying this metric to our planet pairs suggests that the outer four planets are within $\sim$30\% of instability. This analytic stability requirement is another indication, beyond the $n$-body integrations mentioned above, that the planets must have eccentricities less than 0.1-0.2 in order to remain dynamically stable.

Along similar lines, it seems quite unlikely that there are any intermediate planets between Kepler-80d and Kepler-80c as even planets with a fraction of the Earth's mass would reduce the separation between planets and thereby shift the dynamical distance estimates into the unstable regime. Such intermediate planets are also unlikely to avoid transiting (see Brakensiek \& Ragozzine 2016, in press), although planets smaller than $\sim$0.8 Earth radii would likely fall below the detection threshold, even if they were transiting. 

Overall, there seems to be little reason to worry that the inferred masses are inconsistent with dynamical stability; however, an additional long-term perturbation not considered in the above analyses is tidal damping. Very rough estimates of the tidal damping timescales based on a variety of assumptions, and ignoring multi-planet interactions, show that none of the planets are significantly affected by direct semi-major axis decay. Although they are relatively close to their parent star, their small masses raise a paltry tidal bulge. Kepler-80f should be strongly affected by eccentricity damping tides and, assuming terrestrial values for the tidal quality factor $Q$, even Kepler-80d and Kepler-80e could be affected by eccentricity tides (see discussion in $\S$\ref{formation} below). 

Secular excitation greatly complicates this conclusion. In particular, the innermost planet normally would not lose much orbital energy by damping its eccentricity, but if this eccentricity is continually excited by the outer planets, significant orbital decay is likely. It seems quite plausible that the significant separation of the innermost planet is partially due to tidal decay preferentially acting on it, similar to the mechanism proposed by \citet{2014MNRAS.443.1451L} and \citet{2015MNRAS.448.1044H}. In any case, the large innermost period ratio is consistent with the general trend
in all \emph{Kepler} systems seen by \citet{2013ApJ...774L..12S}. We return to the consideration of tides (or other dissipation mechanisms) in investigating the formation of Kepler-80 in $\S$\ref{formation}. 

\subsection{Three-body resonances} \label{3br}

The orbital architecture of Kepler-80 is rare among known systems. The outer four planets are in tight interlocking three-body (mean motion) resonances, meaning that the middle three planets (d, e, and b) and the outer three planets (e, b, and c) are each in three-body resonances. 

Three-body resonances are configurations with resonant angles gives by $\phi = p \lambda_1 - (p+q) \lambda_2 + q \lambda_3$, where $\lambda \equiv \Omega + \omega + M$ is the standard mean longitude \citep{2000ssd..book.....M,2010exop.book..217F}. This equation assumes zeroth-order three-body resonances which are by far the strongest in the case of small eccentricities \citep{2016Icar..274...83G}. This commensurability in periods (since $\frac{2\pi}{P} = n \approx \dot{{\lambda}}$, where $n$ is the mean motion) creates a repeating geometrical configuration of three-planets.  When dynamical interactions cause the resonant angle $\phi$ to librate (according to the pendulum equation), we consider this commensurability to be a \emph{bona fide} three-body resonance as it is stable to perturbations and therefore dynamically meaningful. The most famous example of a three-body resonance is the Laplace resonance visible among the three inner Galilean moons. Such resonances are also important for chaos in the asteroid belt \citep[e.g.,][]{1998AJ....116.3029N} and have been seen in the Pluto system \citep{2015Natur.522...45S} and in the Gliese 876 exoplanetary system \citep[e.g.,][]{2015AJ....149..167B}. \citet{2014MNRAS.445.3959Q} find that three-body resonances can be comparably important as two-body resonances among the small inner moons of Uranus when the moons are near mean-motion resonances. 

\ik observations span multiple $\sim$191-day conjunction super-periods, but do not cover a full three-body resonance libration cycle. We extend the same $n$-body integrations used to produce the TTV fits to explore the dynamical properties of the Kepler-80 system. Note that these integrations do not include the 1-day Kepler-80f since it is assumed to be dynamically decoupled from the other planets (meaning that long-term perturbations are small compared to the resonant effects of interest here). We checked multiple bootstrap fits from both the circular and restricted eccentricity models to confirm the most important results given in this section, and they all give a consistent story.

\begin{figure*}
\centering
\includegraphics[width=.95\textwidth]{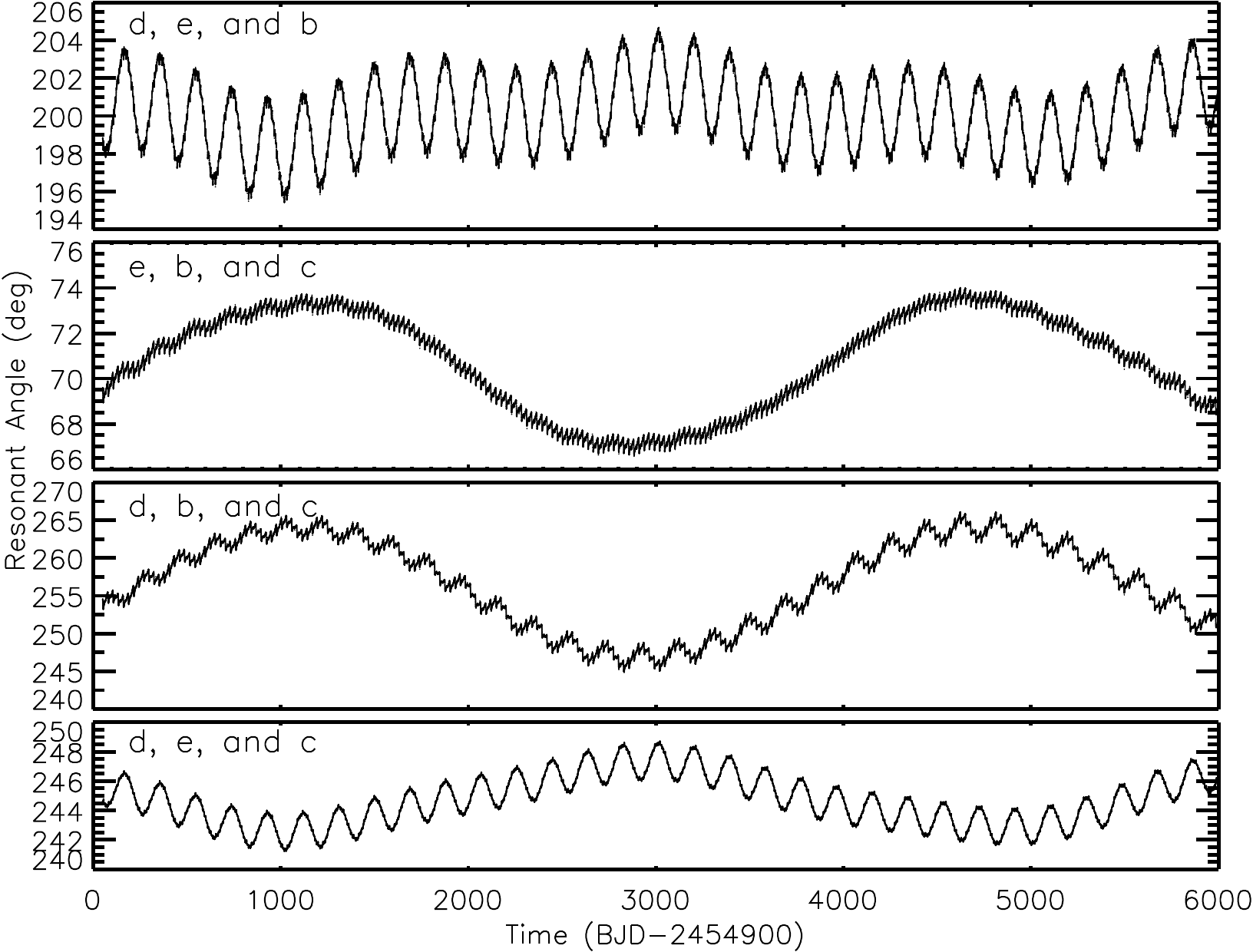}
\caption{Time evolution of the four different three-body resonances seen in Kepler-80. The exact angles are given in the text, e.g., the first plot shows $\phi_1 \equiv 3 \lambda_b - 5 \lambda_e + 2 \lambda_d$ where $\lambda$ is the mean longitude ($\lambda \equiv \Omega + \omega + M$). In Kepler-80, all four of the possible lowest-order three-body resonances are librating with very small $\sim$3$^{\circ}$ libration amplitudes. This four-planet configuration is rare among known planetary-like systems, though three-body (Laplace-like) resonances have been seen. The 191-day conjunction cycle (corresponding to the super-period from the near two-body resonances) is clearly seen. On a longer $\sim$10-year timescale, three-body resonance libration is clearly seen. In both cases, the interlocking resonances produce the same timescale for conjunctions and three-body resonance period. Each of these signatures are seen in the TTV data (Figure \ref{fig:ttvs}) which covers only the first $\sim$1600 days of this plot. Libration centers are shifted from 0 or 180 degrees due to the torque from the planet not in the resonance. This configuration matches the expected result of formation by planetary migration ($\S$\ref{formation}). \label{fig:tbr}}
\end{figure*}

We find that the four possible three-body resonance configurations of Kepler-80 are librating with amplitudes of only a few degrees (see Figure \ref{fig:tbr}), clearly showing that the system is deep in three-body resonances. 

The four-planet commensurability seen in Kepler-80 is due to each pair of planets having almost exactly the same $\sim$191 day ``super-period'' as defined above (Eq. \ref{eq:superperiod}). The TTV signal shows the clear signature due to this 191-day conjunction cycle and an overall quadratic trend that is due to the libration of the three-body resonance which has a period of $\sim$10 years. 

Due to the interlocking nature of the resonances, there are many three-body commensurabilities that are slowly varying in this system. Which three-body resonances are the planets actually in? Two obvious possibilities are the (lowest-order) resonances of adjacent planets: the $\phi_1 \equiv 3 \lambda_b - 5 \lambda_e + 2 \lambda_d$ resonance ($p=2,q=3$) and the $\phi_2 \equiv 2 \lambda_c - 3 \lambda_b + \lambda_e$ resonances ($p=1,q=2$). However, the 1:--2:1 resonance between d, e, and c and the 1:--6:5 resonance between d, c, and b are comparably strong (see Figure \ref{fig:tbr} and discussion below). Many linear combinations of these resonance angles are also librating, which is easy to show mathematically and which we have confirmed by inspection. The arguments of \citet{2014MNRAS.445.3959Q} suggest that these may all contribute to the dynamical evolution and stability of the system. 

Note that the four resonance angles identified in the previous paragraph and shown in Figure \ref{fig:tbr} do not librate around 0 or 180 degrees as theoretically expected for isolated three-body resonances; however, libration around a different center is common in multi-planet systems. It is seen, for example, in Kepler-223/KOI-730 (Mills et al., submitted.) and is caused by a torque from the other, non-resonant, planet that shifts the resonance center from the nominal value. We have confirmed with 1000-year ($\sim$10$^6$ orbits of Kepler-80d) integrations that this dynamical configuration persists with no apparent changes. 

\begin{figure}
\centering
\includegraphics[width=.45\textwidth]{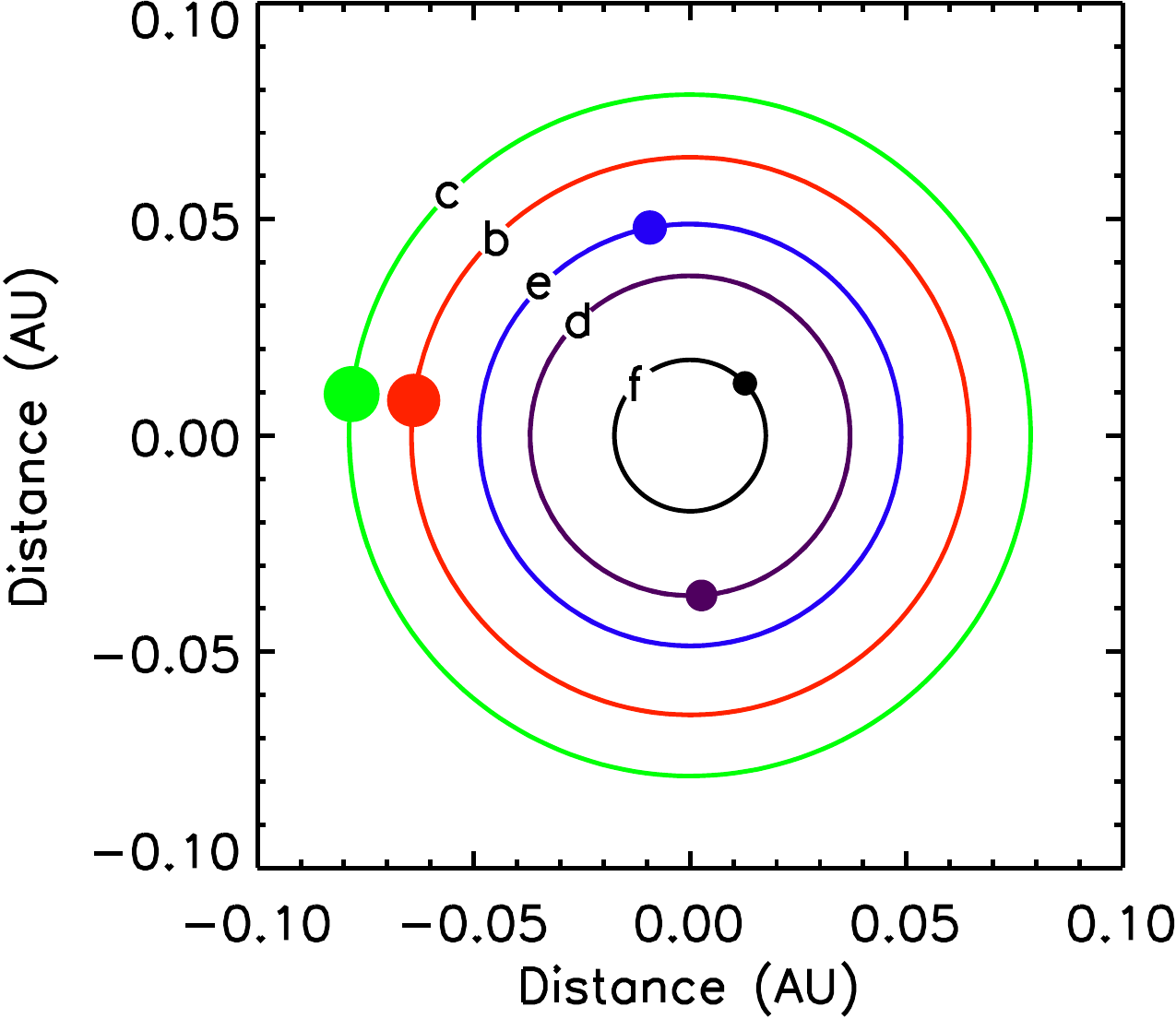}
\caption{A snapshot from our dynamical integration showing the configuration described in the text where a b/c conjunction and a d/e conjunction are nearly 90 degrees apart. This configuration repeats every 191 days due to the four-body commensurability between the orbital periods. The planets are shown to scale with each other, but not to scale with the orbits. Planets d, e, b, and c are shown in yellow, blue, red, and green, respectively. Planet f is shown in black to scale of the others, but as it is not resonant with the others, its position at this time with respect to the other planets is only illustrative.
\label{fig:weirdconj}}
\end{figure}

Through inspection of the times and locations of planets relative to one another at times of conjunctions additional insight can be gained into this unusual dynamical configuration. 
The phases of the planets are such that there is never a triple or quadruple conjunction with three planets aligned, which is likely a resonant protection mechanism that helps ensure long-term stability. There is a time when d and b are aligned and e is anti-aligned, which has similarity to the configuration of the Galilean satellites, and which repeats with a period of 191 days. Another interesting configuration occurs when the outer two planets (b/c) have a conjunction: at this time the inner two planets (d/e) are anti-aligned and the b/c and d/e conjunction lines are nearly 90 degrees from one another. This configuration is shown in Figure \ref{fig:weirdconj}. 

%Through inspection of the times and locations of planets relative to one another at times of conjunctions additional insight can be gained into this unusual dynamical configuration. 
%\textbf{The three different animations of Figure \ref{fig:gifs} show all planets when a pair is in conjunction; from these, it is clear to see the three- and four-body resonances, and the decoupled inner planet.} The phases of the planets are such that there is never a triple or quadruple conjunction with three planets aligned, which is likely a resonant protection mechanism that helps ensure long-term stability. There is a time when d and b are aligned and e is anti-aligned, which has similarity to the configuration of the Galilean satellites, and which repeats with a period of 191 days. Another interesting configuration occurs when the outer two planets (b/c) have a conjunction: at this time the inner two planets (d/e) are anti-aligned and the b/c and d/e conjunction lines are nearly 90 degrees from one another. This configuration is shown in Figure \ref{fig:weirdconj}. 

%\begin{figure}
%\centering
%\includegraphics[width=.33\textwidth]{f6a.eps}
%\includegraphics[width=.33\textwidth]{f6b.eps}
%\includegraphics[width=.33\textwidth]{f6c.eps}
%\caption{Stroboscopic animations available online show the position of all the planets at the times of conjunction of neighboring planets clearly demonstrate the four-body commensurabilities that arise from the interlocking three-body resonances. Planets d, e, b, and c are shown in purple, blue, red, and green, respectively. Planet f is shown in yellow to scale of the others, but as it is not resonant with the others, its position at this time with respect to the other planets is only illustrative.
%\label{fig:gifs}}
%\end{figure}

The small libration amplitudes of Kepler-80's three-body resonances place strong constraints on the formation and past dynamical history of this system, which we now discuss in greater detail.
%%%%%%%%%%%%%%%%%%%%%%%%%%%%%%%%%%%%%%%%%%%%%%%%%%%%%%%%%%%%%%%%%%%

\section{Formation of the Kepler-80 System} \label{formation}

\subsection{Migration Simulations}

Here we describe a scenario, illustrated by a simple numerical implementation, that may have been the evolutionary route that established Kepler-80's three-body resonances.  We propose that convergent migration in the protoplanetary disk placed the outer four planets (d,e,b,c) into a chain of two-body resonances, after which tidal dissipation spread them out of the resonance (e.g., \citealt{2013Batygin}).  It has been noted \citep{2010Papaloizou, 2015IJAsB..14..291P, 2016MNRAS.455L.104G} that such a scenario would likely maintain specific sets of ratios of orbital periods, which dynamically enforces 3-body resonances. 

To model this hypothesis for Kepler-80, we begin by arguing that we can ignore the innermost Kepler-80f and only simulate the outer four planets. One effect of Kepler-80f on the resonances of the outer planets is to provides an effective stellar quadrupole  ($J_2 \simeq 2 \x 10^{-4}$) which would split resonances and can lead to chaos and other interesting effects \citep[][]{1989Icar...78...63T,1990Icar...85..444M}. However, we estimate that the resonance splitting is probably much smaller than the resonance widths and thus not important at its current location. In the past, it is possible that Kepler-80f was originally part of the multi-resonant chain but broke free and was pulled inward by tides. This process might have initially affected the other resonances, but if the migration continues to be strong well after the decoupling of Kepler-80f, it may not affect the final outcome.

Our simulations thus neglected the inner planet and simulated the outer four, choosing each mass as $5 M_{\oplus}$ and a stellar mass of $0.730 M_\odot$. The initial (non-resonant) periods were chosen as 3.1, 4.7, 7.3, and 9.9 days --- slightly more spread from the resonances than the observed system is --- and the orbits were circular.   We followed their Newtonian N-body dynamics using an 8th/9th order Prince-Dormand integration method from the GNU Scientific Library.  In addition, we implemented dissipation with a very simple algorithm, which applies a force to dampen the radial and tangential velocities of individual planets with respect to the host star \citep{2008Thommes}.  To simulate disk migration, we damp the semi-major axis ($e$-folding timescale $10^7$~days) and eccentricity (e-folding timescale $10^5$~days) of the outermost planet (planet c).  It captured the other planets sequentially into the resonances, and we turn this force off at time $5\times10^6$~days$= 1.37\times10^4$~years.  Eccentricity damping (with a timescale $10^5$~days) is applied to the innermost planet d. Tidal evolution is a very strong function of distance, so direct tidal evolution of the other planets was not included. Their evolution is due only to resonant coupling with planet d. 

\begin{figure*} 
\includegraphics[scale=0.62]{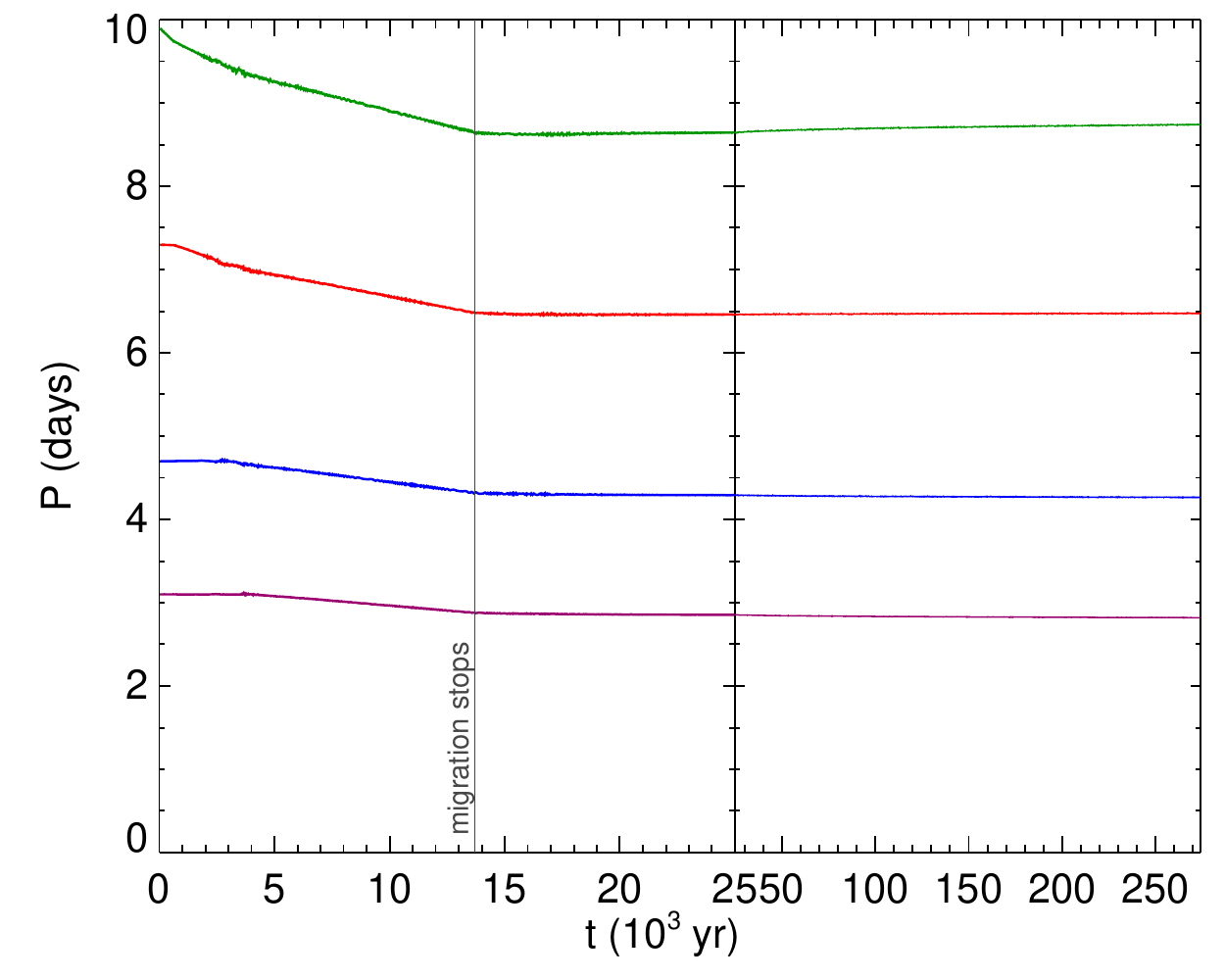}
\includegraphics[scale=0.6]{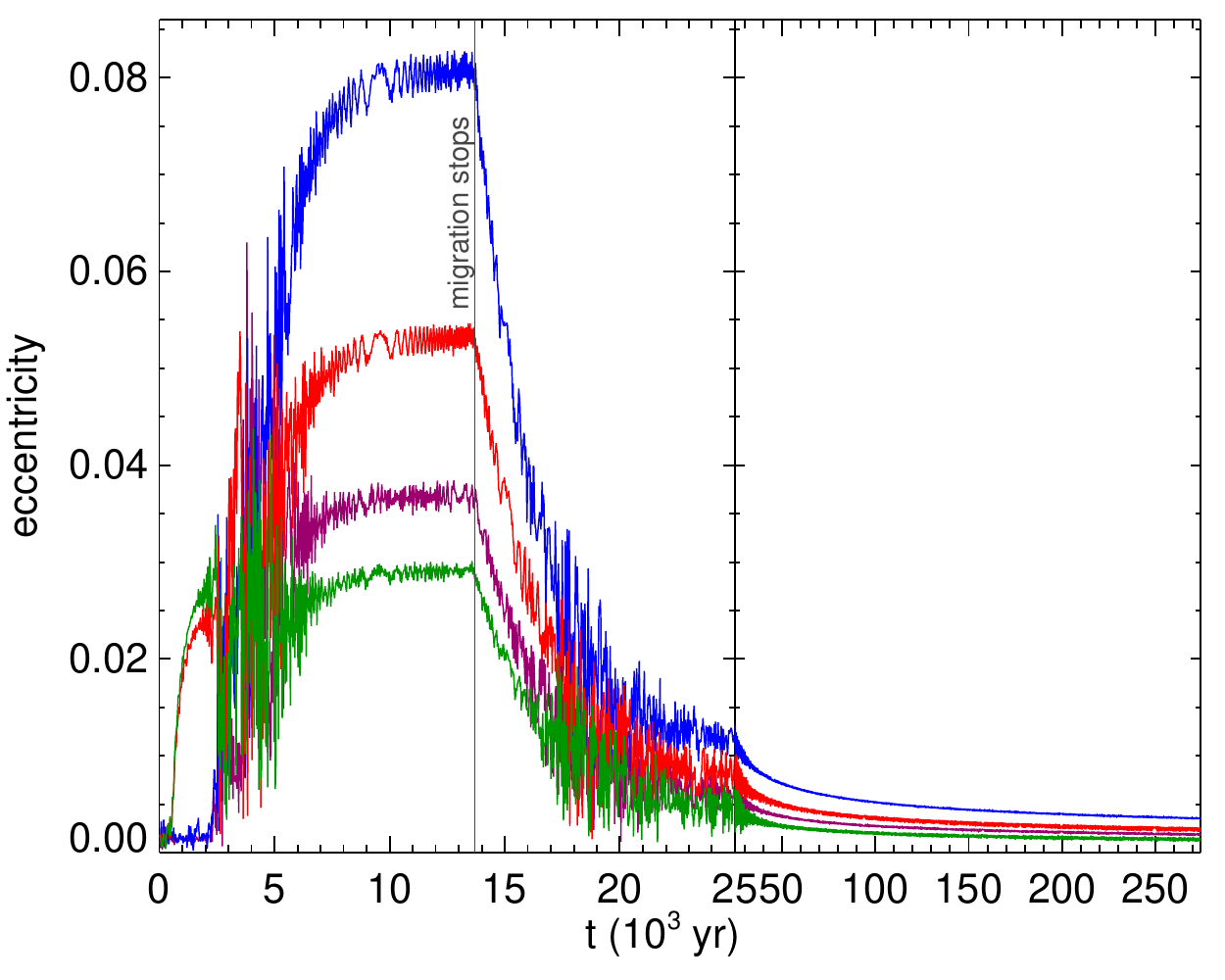}
\includegraphics[scale=0.62]{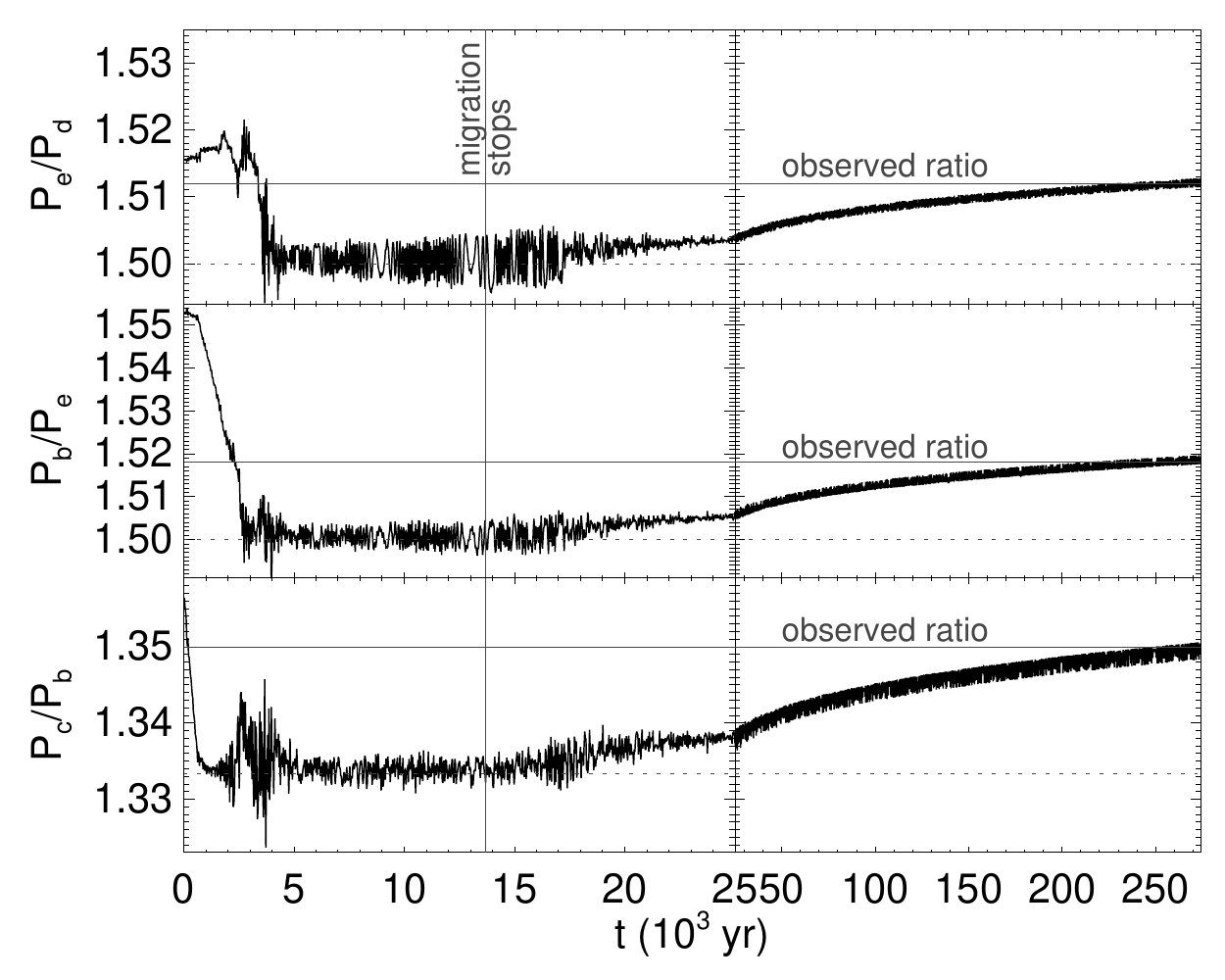}
\includegraphics[scale=0.62]{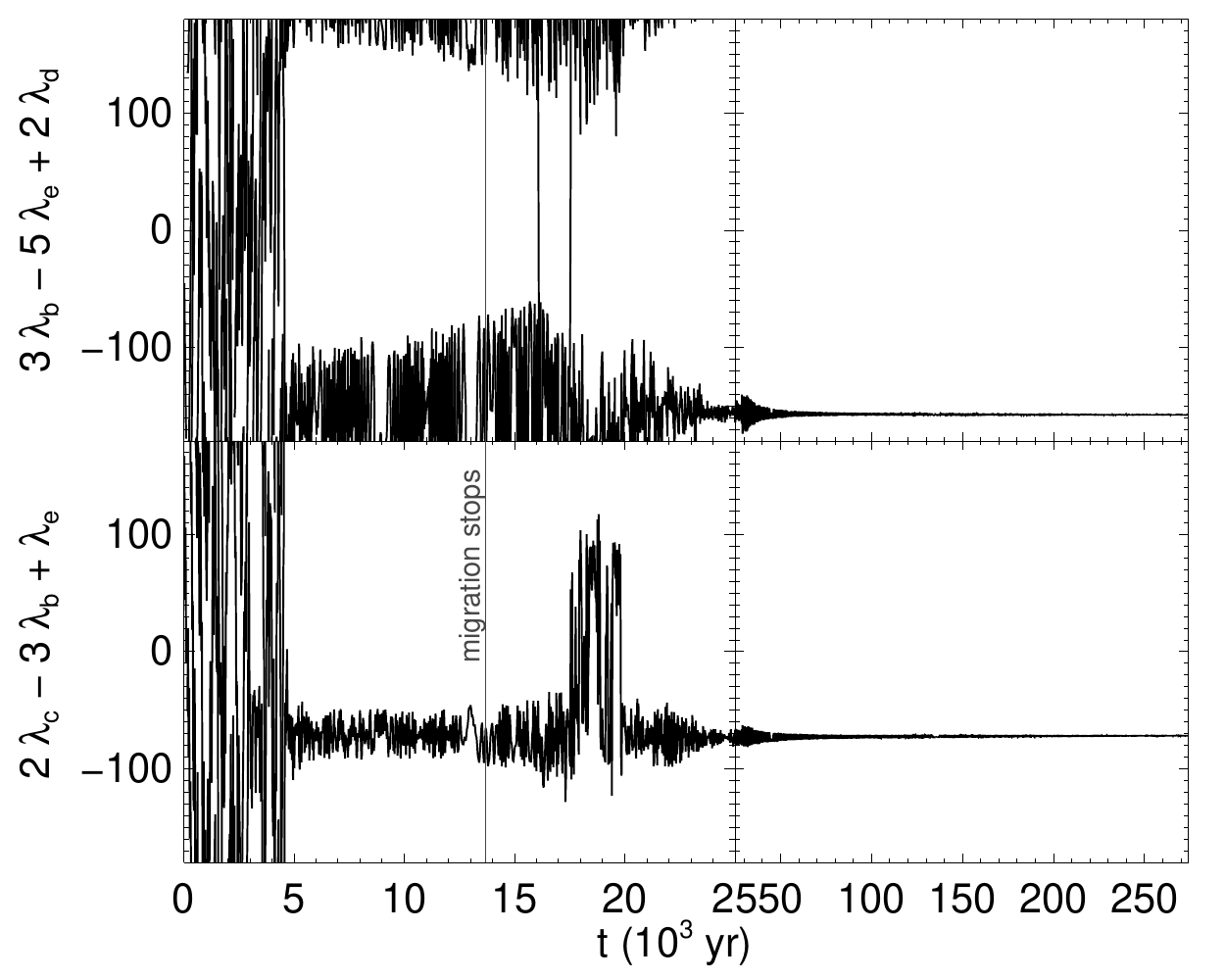}
\caption{Orbital elements, ratios, and critical angles from a numerical experiment designed to reproduce the architecture of the outer four planets of Kepler-80 (d - purple, e - blue, b - red, and c - green, going out). Each panel is horizontally split into two, corresponding to two different timescales, that of type-I disk migration and that of tidal damping, though both are dramatically sped up relative to what they would be in reality. A force that damps the semi-major axis and eccentricity of the outermost planet is active until the vertical line labelled ``migration stops'' in each figure.  In the period-ratio panels, the solid horizontal lines corresponds to the currently-observed period ratios. The libration in the three-body resonances with very small amplitudes is clearly reproduced. Not shown are two-body resonance angles which also librate in this simulation, though the data analysis weakly indicates that two-body resonances are not librating. \label{fig:simulation} }
\end{figure*}

In Figure \ref{fig:simulation}, we show the periods, period ratios, eccentricities, and resonant angles of this simulation.  The orbital period ratios spread out to their observed values, specifically, their observed ratios.  The Laplace resonances were established during migration, and damped further as the inner planet's eccentricity was tidally damped.  We plot the same consecutive-three-body resonance angles ($\phi_1$ and $\phi_2$) shown in the top two subpanels of figure \ref{fig:tbr}.  At the end of the simulation, the resonant libration full amplitudes were  $\sim 1.0^\circ$ and $~1.4^\circ$, on an 8.5 year timescale.  This small libration amplitude seems to match well with the observations (compare Figure \ref{fig:tbr}).  The libration center of $\phi_1$ ($202.5^\circ$) is close to the observed value ($\sim 198^\circ$), whereas the libration center of $\phi_2$ ($-72.0^\circ$) matches in magnitude, with the observed value ($\sim 72.5^\circ$). This is an excellent match considering that there was minimal tuning of orbital parameters beyond starting the planets near their present locations. 

To compare this simulation with reality, the timescales should be lengthened; damping timescales were accelerated to perform the integration more quickly, though slowly enough to perturb the resonances only adiabatically, thereby retaining the correct dynamical character \citep[e.g.,][]{2015IJAsB..14..291P}.  Once the four planets achieved resonant lock, their joint migration changed each of their semi-major axes with a timescale $a / |\dot{a}| \approx 1.8 \times 10^5$~yr.  This corresponds to a factor of about 10 shorter than typical disk lifetimes. We suspect the semi-major axis and eccentricity damping timescales, applied to the outer planet, should be lengthened by this approximate factor.  For tides, using eqn. 5 of \cite{2010Papaloizou}, we find that our eccentricity-damping timescale corresponds to the modified tidal quality factor $Q'=0.126$, which is physically impossible.  A more appropriate long-term time-averaged value for rocky planets is $\sim 10^{2-3}$, meaning the timescales of dissipative divergence seen in the simulation should be lengthened by a factor of $800-8000$, putting the age of the observed systems (given the period ratios) at roughly $1$~Gyr, which is physically reasonable ($\S$\ref{stellar props}). Note that this tidal model is meant to be a reasonable time-averaged approximation to a more advanced and geophysically motivated tidal model \citep[e.g.,][]{2009ApJ...707.1000H, 2014ApJ...795....6E} which is beyond the scope of this work. 

It has been argued that dissipative divergence does not naturally explain many of the offsets from resonance seen in the larger population of \emph{Kepler} planet pairs because the required damping timescales would be unreasonably short \citep{2013Lee} or their free eccentricities would be excessively damped \citep{2012Lithwick}, below the level that is indicated by transit timing in some systems.  In the case of Kepler-80, however, this exercise shows that tidal damping can reasonably explain the shift from resonance. This success may be due to Kepler-80's small semi-major axes compared to most other near-resonant pairs, so that the tidal timescales are reasonably short. 

We also performed a simulation where we stop the inward migration before the two-body resonances are formed. This still leads to a configuration where the observed period ratios are achieved and the system is librating in the two-body and three-body resonances. In this case, the amplitude of the three-body resonance libration is much larger than is seen in the observed data, which suggests that further work may be able to rule out this formation scenario

We consider this model a very non-trivial success in explaining the current architecture of Kepler-80.  Making a more physical model, within which the disk parameters and tidal damping parameters can be constrained, we leave for future work. 

\subsection{Two-Body Resonance Angle Libration}

The successful reproduction of three-body resonance libration encourages us to examine two-body resonances. The orbital periods of the outer four planets are somewhat close to the 3:2, 3:2, and 4:3 mean-motion resonances (moving outwards from the star). In this regard, the Kepler-80 period ratios are similar to the large population of STIPs with planet pairs wide of two-body resonance. We note here that these systems may not technically be in resonance (because of the disappearance of the separatrix) but are expected to still have librating two-body resonance angles \citep[e.g.,][]{2013AJ....145....1B, 2012A&A...546A..71D}. 

Investigation shows that none of the relevant two-body resonance angles are librating in any of our TTV fits, including the 60 bootstrapping fits that explore the parameter space.  This is a surprising result, since other analyses \citep[][]{2010Papaloizou,2013AJ....145....1B, 2015IJAsB..14..291P,2016MNRAS.455L.104G,2016Icar..274...83G} propose that three-body resonances in systems like Kepler-80 would always be accompanied by two-body resonances. Indeed, the migration simulations performed above also produce systems where each of the six possible two-body resonance angles is librating with small amplitude. Furthermore, two-body resonance libration is seen in the similar four-planet Kepler-223 system, though that system is much closer to small integer period ratios than the Kepler-80 planets. Finally, from a theoretical standpoint, it is difficult to conceive a process that would disrupt the two-body resonance angle librations while preserving the three-body resonance, which are generally more fragile and sensitive to perturbations. 

Unlike three-body resonances, two-body resonance angles must include the argument of periapse associated with the eccentricity of one of the two bodies. Given that we cannot reliably recover eccentricities from our TTV analysis ($\S$\ref{validation}), we attempted to determine whether the lack of two-body resonance angle libration was due to insufficient TTV data. We used the successful migration simulation above to produce four-years of TTV measurements similar to our Kepler observations. These data were produced based on the system just beyond time $t=25000$ years, when the ``disk migration'' phase was complete. Though eccentricity damping of the innermost planet is still active, it would have a completely insignificant effect on the TTVs over just four years. We assigned uncertainties and added noise using the uncertainties in our Kepler data (Table \ref{table:scdata}), creating a fake dataset which was then fit using the restricted eccentricity model from $\S$ \ref{reccfit}. 

Our fits showed clear libration of three-body resonance, but no libration of two-body resonance angles. This system is known to show tight libration of two-body resonance angles, but even shrinking the TTV error bars by a factor of $\sim$100 only produced hints of two-body libration. We considered other models (circular, unrestricted eccentricities) -- though not the full suite of analyses presented in $\S$ \ref{validation} -- and never found two-body libration. 

We conclude that identification of two-body resonance angle libration is beyond the capability of the present data and that systems that are actually in resonance will appear to show no libration when fit with our technique. Therefore, the true Kepler-80 planets could easily be librating in the two-body resonances and our formation simulation accurately reproduces the dynamical configuration of Kepler-80 to the limit of our observations. Our success with matching the properties of this system suggests that our simulation approximates the actual dynamical history of Kepler-80.

\subsection{Comparison to Similar Systems}

Two well-studied systems show dynamical similarities to the Kepler-80 system of three-body resonances. Kepler-60 has three-planets that are also in or near the three-body resonance. \citet{2016MNRAS.455L.104G} perform a TTV analysis of Kepler-60 and find that the three-body resonance angle is librating; however, this analysis did not account for possible overfitting of the eccentricities (see $\S$\ref{validation} above), and another analysis by \citet{2016ApJ...820...39J} finds that only $\sim$80\% of the TTV fits show libration of the Laplace angle.  \citet{2016MNRAS.455L.104G} show that the TTVs cannot distinguish whether the system is also in two-body resonances, as we find for Kepler-80. The work of \citet{2015IJAsB..14..291P} on the Kepler-60 system would indicate that, if two-body resonances were originally present, they could be preserved as the system evolved to its current state due to tides (see also $\S$\ref{formation}). 

Kepler-223/KOI-730 is a four-planet system that also has two interlocking three-body resonances studied extensively by \citet{2016Natur.533..509M}. Its planets are in two-body resonances with each other forming a multi-resonant chain that also supports the three and four-body commensurabilities seen in Kepler-80. However, the dynamical distance to two-body resonance in the Kepler-80 system is much greater than the Kepler-223 system.

We note here that there are likely connections between Kepler-80, Kepler-60, Kepler-223, and the many near-resonant pairs seen by \emph{Kepler}. These systems may be consequences of similar formation processes with present-day differences due to an evolutionary sequence and/or different key properties yet to be identified. Combining our understanding of these systems could allow us to infer more about how the formation and evolution of exoplanetary systems. In this regard, it is exciting to see that Kepler-80 has the best TTV measurements of these three systems, due to higher signal-to-noise transits and shorter dynamical timescale. 

\section{Conclusions} \label{conclusion}

\ik has provided us with a wealth of data on the architectures of planetary systems. Herein, we investigated the dynamically intriguing Kepler-80 system (planets f, d, e, b, and c in order of period) and came to several interesting conclusions. 

A self-consistent dynamical analysis of the system, using TTV fitting under the assumption of restricted eccentricities, inferred masses for the outer four planets (d, e, b, and c) of $6.75^{+0.69}_{-0.51}$, $4.13^{+0.81}_{-0.95}$, $6.93^{+1.05}_{-0.70}$, and $6.74^{+1.23}_{-0.86}$ Earth masses, respectively. The choice to restrict eccentricities to small values resulted from extensive testing of our fitting technique that showed TTV fits using eccentric models infer accurate mass estimates but inaccurate eccentricities and apsidal angles. Further tests showed that we cannot infer two-body resonance angle libration with Kepler-80 TTVs.

Although all four planets have very similar masses, planets d and e are terrestrial and planets b and c have $\sim$2\% (by mass) H/He envelopes assuming Earth-like cores. Their orbits are similar and models suggest that photo-evaporation would have removed $\sim$1\% H/He from all four planets. Though simulations suggest the system has been affected by planetary tides, we did not consider the effect of dissipation on the atmospheric history of the planets. It is unusual to have four well-measured densities in the same system and future comparative planetology may constrain the formation and evolution of their atmospheres.

Kepler-80 is very interesting dynamically. The system appears to be long-term stable as long as eccentricities are below $\sim$0.2. The outer four planets in Kepler-80 are in a dynamically rare configuration, with multiple three-body resonances librating with only $\sim$3$^{\circ}$ amplitude. This architecture is the natural result of migration simulations, described herein, where the four outer planets were in a resonant chain and a dissipative forces pushed them wide of nominal two-body resonance locations (while retaining two-body resonance angle libration) and deep into three-body resonances. Kepler-80 should thus play an important constraint on the formulation and evolution of STIPS.

Many of these conclusions are fruitful starting points for additional study. To assist in the future observational efforts, we extrapolate our restricted eccentricity bootstrap models $\sim$15 years into the future and provide the transit times and estimated uncertainties in Table \ref{table:ttestimates}. Four years of high-precision coverage from \ik has maintained the uncertainty in near-term (e.g., 2016) transit times to about 10 minutes for each planet; a TT measurement more precise than this will be required to significantly improve the model. For transits with a depth of 0.5-1.6 millimagnitudes and a duration of 2 hours on a $V \simeq 15.2$ magnitude star, useful TTV measurements will require space-based observations with large aperture telescopes. Neither TESS \citep{2014SPIE.9143E..20R} nor CHEOPS \citep{2013EPJWC..4703005B} will have sufficient precision. At the estimated time of PLATO observations of the \ik field \citep{2014ExA....38..249R}, the TT uncertainty for the planets will have grown to about 30 minutes, which may be detectable. Note that these are statistical uncertainty estimates that do not include potential sources of systematic error. 

\begin{deluxetable*}{lcccc}
\tabletypesize{\footnotesize}
\tablecolumns{5}
\tablewidth{0pt}
\tablecaption{Transit Time Predictions\label{table:ttestimates}}
\tablehead{
\colhead{Planet} &
\colhead{Transit No.} &
\colhead{Transit Time (BJD)} &
\colhead{$\sigma^+$} &
\colhead{$\sigma^-$}
}

\startdata
Kepler-80d	&	0	&	2455695.130 &	0.003055367	&	0.004165690	\\
Kepler-80d	&	1	&	2455698.202	&	0.003098780 &	0.004228180	\\
Kepler-80d	&	2	&	2455701.275	&	0.003295472	&	0.004198860	\\
Kepler-80d	&	3	&	2455704.347	&	0.003348749	&	0.004229972	\\
Kepler-80d	&	4	&	2455707.419	&	0.003356507	&	0.004354759	\\
Kepler-80d	&	5	&	2455710.492	&	0.003385405	&	0.004236598	\\
Kepler-80d	&	6	&	2455713.564	&	0.003373803	&	0.004251005	\\
Kepler-80d	&	7	&	2455716.637	&	0.003396669	&	0.004194434	\\
Kepler-80d	&	8	&	2455719.709	&	0.003519851	&	0.003935169	\\
Kepler-80d	&	9	&	2455722.782	&	0.003535066	&	0.003922528	\\
Kepler-80d	&	10	&	2455725.854	&	0.003600991	&	0.003827943	\\

\enddata
\tablecomments{Predictions of future transit times from our integrations through 2025. These predictions were made using our eccentric bootstrapping models. The columns, from left to right, are: the planet's Kepler name, the transit number (where transit 0 indicates the first transit after the epoch of BJD 2454693), the transit time (BJD), the upper uncertainty on the transit time taken to include the 84$^{th}$ percentile ($\sigma^+$), and the lower uncertainty on the transit time taken to include the 16$^{th}$ percentile ($\sigma^-$). Uncertainties are given in units of days and are about 10 minutes in the near term (2016) and grow to about 30 minutes by 2025. Table \ref{table:ttestimates} is published in its entirety in the electronic edition of the \emph{Astrophysical Journal}. A portion is shown here for guidance regarding its form and content.}
\end{deluxetable*}

A full photodynamical model of Kepler-80 (with stellar parameters updated after GAIA) is a worthwhile endeavor to somewhat improve mass and eccentricity estimates and uncertainties as well as the covariances. For example, we did not use the known durations to constrain the system, which might help to constrain the eccentricities in a less artificial way. Combination with a Bayesian technique would be particularly powerful, as it would be much less susceptible to overfitting, an issue which plagued our inference of eccentricities, apsidal angles, and two-body resonance libration. Additional investigation into the meaning and origin of the three-body resonances might provide interesting constraints of the formation of this system (e.g., damping timescales), which may be broadly applicable to other STIPs. 

Kepler-80 has proven to be an information-rich multi-transiting system, and we hope that future endeavors will continue to provide insight into this system with implications for the formation and evolution of planetary systems.

\begin{acknowledgments}

We thank the \ik Team for their outstanding work. We thank many for their comments, suggestions, and contributions including Bill Cochran, Debra Fischer, Daniel Jontof-Hutter, Geoff Marcy, Tim Morton, Sam Quinn, Victor Sallee, Ren\'e Hille, and $Kepler$'s TTV/Multis Working Group. 

The authors wish to extend special thanks to those of Hawai'ian ancestry on whose sacred mountain of Mauna Kea we are privileged to be guests. Without their generous hospitality, the Keck observations presented herein would not have been possible.

\end{acknowledgments}

\bibliographystyle{apj}
\bibliography{ms}

\end{document}